\documentclass{article}
\usepackage{graphicx}
\usepackage{latexsym}
\usepackage{amsmath}
\usepackage{amssymb}
\usepackage{amsfonts}
\usepackage{amsxtra}
\usepackage{cite}
\usepackage{xr}

\allowdisplaybreaks
\newcommand{\ver}[1]{\widehat{#1}}
\newcommand{\verr}[1]{\ver{r}^{\,#1}}
\newcommand{\verrp}[1]{\widehat{r}\,'^{#1}}
\newcommand{\verp}[1]{\ver{p}^{\,#1}}
\newcommand{\verpp}[1]{\widehat{p}\,'^{#1}}

\newcommand{\vchiU}[2]{\ver{\chi}_{#1}^{\,#2}}
\newcommand{\veps}[1]{\ver{\varepsilon}^{\,#1}}
\newcommand{\CG}[6]{\langle #1,#2;#3,#4 | #5,#6\rangle}

\renewcommand{\P}{\mathcal{P}}
\newcommand{\lave}{\langle\ell\rangle}
\newcommand{\dl}{\Delta\ell}

\title{Addition theorems for spin spherical harmonics. II Results} 
\author{Antonio O.\ Bouzas \thanks{E-mail:
    abouzas@mda.cinvestav.mx}\\\small Departamento de F\'{\i}sica
  Aplicada, CINVESTAV-IPN \\\small Carretera Antigua a Progreso Km.\
  6, Apdo.\ Postal 73 ``Cordemex''\\\small M\'erida 97310, Yucat\'an,
  M\'exico}
\date{\today}
\begin{document}
\maketitle
\begin{abstract}
  Based on the results of part I, we obtain the general form of the
  addition theorem for spin spherical harmonics and give explicit
  results in the cases involving one spin-$s'$ and one spin-$s$
  spherical harmonics with $s',s=1/2$, 1, 3/2, and $|s'-s|=0$, 1.  We
  obtain also a fully general addition theorem for one scalar and
  one tensor spherical harmonic of arbitrary rank.  A variety of
  bilocal sums of ordinary and spin spherical harmonics are given in
  explicit form, including a general explicit expression for bilocal
  spherical harmonics.
\end{abstract}

\section{Introduction}
\label{sec:intro}

In this paper we study bilocal sums of spin spherical harmonics
\cite{bie85a,var88} (see sect.\ \ref{sec:ssh} below for more
details).  Such expressions occur in the computation of physical
observables in systems undergoing dynamical transitions induced by
rotationally invariant spin-dependent interactions.  Expanding the
states involved in such transitions in a basis of total
angular-momentum eigenfunctions leads to the sums considered here.  In
quantum systems, partial-wave expansions of transition amplitudes, or
other matrix elements, are given by addition theorems for spin
spherical harmonics.  Another example is provided by multipolar
expansions of the classical electromagnetic radiation field
\cite{jackxx} in quadratic observables like energy, energy flow, or
radiated power, which lead in some cases to bilocal sums of vector
spherical harmonics.

Our main motivation for the present paper is the application of its
results to the partial-wave analysis of scattering processes.  In that
context, partial waves satisfying two-body unitarity play an essential
role in the theoretical approach (see \cite{bou09} for a realistic
application to hadronic reactions and an extensive reference list).
Two-body intermediate states with different spins must be included in
the analysis in order to satisfy unitarity, both as genuine asymptotic
states or, in the case of resonances, as an approximation to higher
$n$-body intermediate states.  The addition theorems we consider in
this paper are then needed for a systematic approach to partial-wave
expansions of $S$-matrix elements between states involving particles
with different spins.  For that reason, we expect the results given
below to be of interest also in other fields where quantum scattering
is relevant.  More generally, however, given the ubiquitous
applications of spherical harmonics in all areas of physics as
solutions to the Laplace equation, we hope the addition theorems
discussed here to be of interest by themselves, independently of our
specific motivations.

Our discussion of bilocal sums of spin spherical harmonics in the
following sections is entirely based on results from the first part of
this paper \cite{bou10}, hereafter referred to as I.  We refer to
equations and sections of the first part by prefixing their numbers
with I.  The matrix elements obtained in sect.\ I.\ref{I.sec:proj} are
restated below as bilocal sums of ordinary spherical harmonics.  In
particular, they provide an explicit general expression for bilocal
spherical harmonics \cite{var88}. The results of sect.\
I.\ref{I.sec:proj} also lead to a completely general addition theorem
for, and to some classes of bilocal sums of, one scalar and one
rank-$n$ tensor spherical harmonic for any $n>0$.  Using the results
of sects.\ I.\ref{I.sec:fac} and I.\ref{I.sec:proj}, we obtain below a
general addition theorem for two spin spherical harmonics with spins
$s',s>0$, and discuss it in completely explicit form in the cases
$s',s \leq 3/2$, $|s'-s|=0,1$.  Our approach can easily be applied to
higher-spin spherical harmonics, although for $s>2$ the size of the
resulting algebraic expressions grows rapidly with $s$.

The paper is organized as follows.  In the next section we briefly
review the definition and main properties of spin-$s$ spherical
harmonics, for any integer or half-integer $s$. In section
\ref{sec:2scalar} we study bilocal sums of ordinary spherical
harmonics, including bilocal spherical harmonics.  In section
\ref{sec:scalspin} we discuss bilocal sums involving one scalar and
one tensor spherical harmonic.  Addition theorems involving two spin
spherical harmonics are considered in sect.\ \ref{sec:spin}.  Finally,
in section \ref{sec:finrem} we offer some final remarks. Throughout
the paper we follow the notation and conventions of appendices
I.\ref{I.sec:notatio} and I.\ref{I.sec:standard}.

\section{Spin spherical harmonics}
\label{sec:ssh}

Spin-$s$ spherical harmonics $Y^{\ell s}_{jj_z}(\verr{}) =
\langle\verr{}, \psi_s | \ell,s,j,j_z\rangle$ \cite{bie85a,var88} are
eigenfunctions of $\vec{L}^{\,2}$, $\vec{S}^{\,2}$, $\vec{J}^{\,2}$,
$J_z$, appropriate for the description of a spin-$s$ particle subject
to spin-dependent central interactions.  Expansion of the state $|
\ell,s,j,j_z\rangle$ in the product basis $| \ell,\ell_z;
s,s_z\rangle$ yields
\begin{equation}
  \label{eq:def}
  Y^{\ell s}_{jj_z}(\verr{}) = \sum_{\ell_z=-\ell}^\ell
  \sum_{s_z=-s}^s \CG{\ell}{\ell_z}{s}{s_z}{j}{j_z} \langle\verr{} |
  \ell,\ell_z\rangle \langle \psi_s | s, s_z\rangle~,
\end{equation}
where $|\verr{}\rangle$ denotes a position eigenstate on
the unit sphere, and $|\psi_s\rangle$ a basis spin state.

For integer spin $n>0$ the spin wave functions are given by the standard
rank-$n$ irreducible tensors defined in sect.\ I.\ref{I.sec:tensor},
$\langle \psi_s | s, s_z\rangle =\veps{i_1\ldots i_s}(s_z)$, $-s\leq
s_z \leq s$, so that
\begin{equation}
  \label{eq:ndef}
\left(Y_{jj_z}^{\ell n}(\verr{})\right)^{i_1\ldots i_n} =
\sum_{s_z=-n}^{n}\sum_{\ell_z=-\ell}^\ell
\CG{\ell}{\ell_z}{n}{s_z}{j}{j_z} Y_{\ell\ell_z}(\verr{})
\veps{i_1\ldots i_n}(s_z)~.  
\end{equation}
For $n=1$ this equality reduces to the definition of vector spherical
harmonics familiar from textbooks \cite{gal90,jackxx}.  For $n>1$,
spin-$n$ spherical harmonics are totally symmetric and traceless, as
shown in appendix I.\ref{I.sec:tensor}.  From that appendix we also
obtain,
\begin{equation}
  \label{eq:nrel}
  \begin{gathered}
  {\left(Y^{\ell n}_{j j_z}(\verr{})\right)^{i_1\ldots i_n}}^* =
  (-1)^{j_z} (-1)^{\ell+n-j} \left(Y^{\ell n}_{j
      (-j_z)}(\verr{})\right)^{i_1\ldots i_n}~, \\
  \int d^2p\, {\left(Y_{jj_z}^{\ell n}(\verp{})\right)^{i_1\ldots i_n}}^*
  \left(Y_{j'j'_z}^{\ell' n}(\verp{})\right)^{i_1\ldots i_n} =
  \delta_{\ell'\ell} \delta_{j'_z j_z} \delta_{j'j}~, \\
  \sum_{j,\ell=0}^\infty \sum_{j_z=-j}^j \left(Y_{jj_z}^{\ell
      n}(\verr{})\right)^{i_1\ldots i_n} {\left(Y_{jj_z}^{\ell
      n}(\verrp{})\right)^{j_1\ldots j_n}}^* =
  \delta\left(\verr{}-\verrp{}\right) X^{i_1\ldots i_n;j_1\ldots j_n},  
  \end{gathered}
\end{equation}
where in the last equality $X^{i_1\ldots i_n;j_1\ldots j_n}$ is the
orthogonal projector onto the subspace of irreducible 
tensors of $\mathbb{C}^{3\times \ldots \times 3}$ defined in
(I.\ref{I.eq:tensorcomplete}). 

For half-integer spin, $s=n+1/2$, $n\geq 0$, the spin wave functions
are given by the standard spinors defined in sect.\
I.\ref{I.sec:spinor}, $\langle \psi_{n+1/2} | n+1/2, s_z\rangle
=\vchiU{A}{i_1\ldots i_n}(s_z)$, which leads to spin-$(n+1/2)$
spherical harmonics defined as
\begin{equation}
  \label{eq:3halfdef}
\left(Y_{jj_z}^{\ell (n+\frac{1}{2})}(\verr{})\right)_A^{
i_1\ldots i_n} =
\sum_{s_z=-n-1/2}^{n+1/2}\sum_{\ell_z=-\ell}^\ell
\CG{\ell}{\ell_z}{n+1/2}{s_z}{j}{j_z} Y_{\ell\ell_z}(\verr{}) 
\vchiU{A}{i_1\ldots i_n}(s_z)~.
\end{equation}
For $n=0$ we have the definition of spin-1/2 spherical harmonics,
possessing two independent complex components.  For $n>0$, from
appendix (I.\ref{I.sec:spinor}) we have that 
spin-$(n+1/2)$ spherical harmonics are totally simmetric and traceless
in their tensor indices and satisfy,
\begin{equation}
  \label{eq:nhalfpolar}
\sigma^{i_k}_{AB} \left(Y^{\ell(n+1/2)}_{j j_z}(\verr{})
\right)_{B}^{i_1\ldots i_k\ldots i_n}=0~,
\qquad
1\leq k\leq n~.
\end{equation}
From the properties of spinors in appendix (I.\ref{I.sec:spinor}) we
also obtain,
\begin{equation}
  \label{eq:3halfrel}
  \begin{gathered}
i\sigma^2_{AB} \left( Y^{\ell(n+1/2)}_{j j_z}(\verr{})
\right)^{i_1\ldots i_n}_{B} = (-1)^{\ell+n+1/2-j} (-1)^{1/2+jz}
{\left( Y^{\ell(n+1/2)}_{j (-j_z)}(\verr{})
\right)^{i_1\ldots i_n}_{B}}^*~,
\\
  \int d^2p\, {\left(Y_{jj_z}^{\ell
  (n+\frac{1}{2})}(\verp{})\right)^{i_1\ldots i_n}_A}^* 
  \left(Y_{j'j'_z}^{\ell' (n+\frac{1}{2})}(\verp{})\right)^{
i_1\ldots i_n}_A =
  \delta_{\ell'\ell} \delta_{j'_z j_z} \delta_{j'j}~,
\\
  \sum_{j,\ell=0}^\infty \sum_{j_z=-j}^j \left(Y_{jj_z}^{\ell
   (n+\frac{1}{2})}(\verr{})\right)^{i_1\ldots i_n}_A
{\left(Y_{jj_z}^{\ell (n+\frac{1}{2})}(\verrp{})\right)^{j_1\ldots
    j_n}_B}^* = \delta\left(\verr{}-\verrp{}\right) X^{i_1\ldots i_n;
j_1\ldots j_n}_{AB}~,
  \end{gathered}
\end{equation}
where $X^{i_1\ldots i_n;j_1\ldots j_n}_{AB}$ is the orthogonal
projector onto the spin-$(n+1/2)$ subspace of $\mathbb{C}^{3\times
  \ldots\times3\times 2}$ defined in (I.\ref{I.eq:nspincompl}).

\section{Bilocal sums of  two scalar spherical harmonics}
\label{sec:2scalar}

The matrix elements of the angular-momentum projector operators with
tensor products of $\vec{L}$ and $\verr{}$ obtained in section
I.\ref{I.sec:proj} constitute addition theorems, or weighted sums, for
two scalar spherical harmonics.  In this section we summarize the main
results.

From the definition (I.\ref{I.eq:proj}) of $\P_\ell$ we have,
\begin{equation}
  \label{eq:mix1}
  \begin{split}
  \sum_{\ell'_z=-\ell_n}^{\ell_n}\sum_{\ell_z=-\ell}^\ell Y_{\ell_n \ell'_z}(\verpp{}) Y_{\ell
    \ell_z}(\verp{})^* \langle\ell_n,\ell'_z | \verr{\{i_1}\ldots
  \verr{i_n} L^{j_1}\ldots L^{j_t\}_0} &| \ell, \ell_z\rangle = \\
  &=\langle\verpp{} | \P_{\ell_n} \verr{\{i_1}\ldots
  \verr{i_n} L^{j_1}\ldots L^{j_t\}_0} \P_\ell |\verp{}\rangle~.
  \end{split}
\end{equation}
The matrix element on the r.h.s.\ of (\ref{eq:mix1}) is given
explicitly for any $\ell\ge0$ by (I.\ref{I.eq:mix4}) and
(I.\ref{I.eq:mix5}) for $n>0$, $t>0$, by (I.\ref{I.eq:genL}) and
(I.\ref{I.eq:Ls}) for $n=0$, $t>0$, and by (I.\ref{I.eq:rmat3}) and
(I.\ref{I.eq:rmat4}) for $n>0$, $t=0$.  Notice that, in the last case,
the matrix element in (\ref{eq:mix1}) involves an additional
symmetrization with respect to (I.\ref{I.eq:rmat3}). The equalities
(I.\ref{I.eq:mix4}), (I.\ref{I.eq:genL}) and (I.\ref{I.eq:rmat3})
written in the form (\ref{eq:mix1}) will be needed below in the
derivation of addition theorems for spin spherical harmonics.  By
multiplying (\ref{eq:mix1}) by a standard tensor of rank $n+s$, as
defined in appendix (I.\ref{I.sec:standard}), and applying the
Wigner-Eckart theorem
\cite{edm96,wig59,jud75,bie85a,bie85b,var88,gal90} to the matrix
element on the l.h.s., we obtain the equivalent form
\begin{equation}
  \label{eq:mix2}
  \begin{split}
  \sum_{\ell_z=-\ell}^\ell Y_{\ell_n(\ell_z+m)}(\verpp{}) Y_{\ell
    \ell_z}(\verp{})^*
  &\CG{\ell}{\ell_z}{n+t}{m}{\ell_n}{\ell_z+m} =
  \\ 
  &=\frac{1}{(n+t)!}\veps{h_1\ldots h_nk_1\ldots k_t}(m)
  \frac{\langle\verpp{} | \P_{\ell_n} \verr{\{h_1}\ldots
  \verr{h_n} L^{k_1}\ldots L^{k_t\}_0} \P_\ell |\verp{}\rangle}
  {\langle\ell_n || \veps{a_1\ldots a_nb_1\ldots
    b_t} \verr{a_1}\ldots \verr{a_n} L^{b_1}\ldots L^{b_t} ||
  \ell\rangle}~,
  \end{split}
\end{equation}
where the reduced matrix element in the denominator on the r.h.s.\ is
given by (I.\ref{I.eq:redmat7}) for $n>0$, $t>0$, by
(I.\ref{I.eq:redmat3}) for $n=0$, $t>0$, and by (I.\ref{I.eq:redmat6})
for $n>0$, $t=0$.  Equation (\ref{eq:mix2}) will be recast below in
sect.\ \ref{sec:scalspin} in the form of an addition theorem for one
scalar and one tensor spherical harmonic.

Equations (\ref{eq:mix1}) and (\ref{eq:mix2}) contain a great deal of
information.  First, together with the matrix elements given
explicitly in I, (\ref{eq:mix2}) provides an explicit representation
for bilocal spherical harmonics \cite{var88}, the wave functions
associated to coupled orbital angular momentum states $|\ell, \ell',
j, j_z\rangle$
\begin{equation*}
  \label{eq:mix2aux}
  \begin{split}
\langle \verp{}, \verpp{} | \ell,\ell',n,m\rangle  &= 
\sum_{\ell'_z=-\ell'}^{\ell'} \sum_{\ell_z=-\ell}^{\ell} 
Y_{\ell'\ell'_z}(\verpp{}) Y_{\ell\ell_z}(\verp{})
\langle \ell, \ell_z; \ell',\ell'_z | n,m\rangle\\
&= \frac{(-1)^{n-\ell'}}{n!} \sqrt{\frac{2n+1}{2\ell'+1}}
\veps{h_1\ldots h_n}(m)
  \frac{\langle\verpp{} | \P_{\ell'} \verr{\{h_1}\ldots
  \verr{h_{|\dl|}} L^{h_{|\dl|+1}}\ldots L^{h_n\}_0} \P_\ell
  |\verp{}\rangle} 
  {\langle\ell' || \veps{a_1\ldots a_n} \verr{a_1}\ldots
    \verr{a_{|\dl|}} L^{a_{|\dl|+1}}\ldots L^{a_n} ||
  \ell\rangle}~.
  \end{split}
\end{equation*}
Second, seen as a generic sum of products of two spherical harmonics
weighted by a CG coefficient, (\ref{eq:mix2}) is best encoded in terms
of tensor spherical harmonics, as done in the following section.
Finally, by considering the explicit expression of the CG on its
l.h.s.\ as a function of $\ell_z$, we can interpret (\ref{eq:mix2}) as
a sum of bilocal products of two spherical harmonics weighted by
specific functions.  We discuss several such bilocal sums in the
remainder of this section.

If we set $n=0$, (\ref{eq:mix1}) involves only matrix elements of
$\vec{L}$.  We will discuss here two classes of bilocal sums obtained
from (\ref{eq:mix1}) in that case.  First, we consider moments of
$\ell_z$.  We have
\begin{equation}
  \label{eq:momaux}
  \begin{aligned}
    \langle\ell,\ell_z | L^{\{3}\ldots L^{3\}_0}| \ell,\ell_z\rangle
  &= t!\, \veps{3\ldots3}(0)^* \langle\ell || \veps{j_1\ldots j_t}
  L^{j_1}\ldots L^{j_t} ||
  \ell\rangle \CG{\ell}{\ell_z}{t}{0}{\ell}{\ell_z}\\
  &= \frac{(t!)^2}{2^t (2t-1)!!} \frac{1}{\sqrt{2\ell+1}}
  \sqrt{\frac{(2\ell+t+1)!}{(2\ell-t)!}}
  \CG{\ell}{\ell_z}{t}{0}{\ell}{\ell_z},
  \end{aligned}
\end{equation}
where the second equality follows from (I.\ref{I.eq:tensor2}) applied
to $\veps{3\ldots3}(m)$.  The CG coefficient on the r.h.s.\ of
(\ref{eq:momaux}) is a polynomial of degree $t$ in $\ell_z$, so
substituting (\ref{eq:momaux}) on the l.h.s.\ of (\ref{eq:mix1}) we
obtain the desired moments.  For $t\leq 3$ the irreducible tensors
appearing in the r.h.s.\ of (\ref{eq:mix1}) (see (I.\ref{I.eq:genLa}))
are easily expanded to get, with the notation of (I.\ref{I.eq:genLb}) 
\begin{subequations}
  \label{eq:momres}
\begin{align}
  \label{eq:momresa}
  \sum_{\ell_z=-\ell}^\ell \ell_z Y_{\ell \ell_z}(\verpp{})
  Y_{\ell \ell_z}(\verp{})^*  &= i \frac{2\ell+1}{4\pi} v^3
  P'_\ell(x)~, \\
  \label{eq:momresb}
  \sum_{\ell_z=-\ell}^\ell {\ell_z}^2 Y_{\ell \ell_z}(\verpp{})
  Y_{\ell \ell_z}(\verp{})^*  &= -\frac{2\ell+1}{4\pi} \left(
    \left(v^3\right)^2 P''_\ell(x) + \left( \verp{3}\verpp{3} -
      x\right) P'_\ell(x) \right)~,\\
  \label{eq:momresc}
\sum_{\ell_z=-\ell}^\ell {\ell_z}^3 Y_{\ell \ell_z}(\verpp{})
  Y_{\ell \ell_z}(\verp{})^*  &= 
  \begin{aligned}[t]
   & -i \frac{2\ell+1}{4\pi} v^3 \left(
    \left(v^3\right)^2 P'''_\ell(x) + 3 \left( \verp{3}\verpp{3}
  - x\right) P''_\ell(x) - P'_\ell(x)\rule{0pt}{10pt} \right), 
  \end{aligned}
\end{align}
where the Legendre equation was used to simplify the
right-hand sides.  For $t>3$ the relations among tensors and their
irreducible components become too lengthy, so we do not expand them
\begin{align}
  \label{eq:momresd}
 & \begin{aligned}
\sum_{\ell_z=-\ell}^\ell &{\ell_z}^4 Y_{\ell \ell_z}(\verpp{})
  Y_{\ell \ell_z}(\verp{})^*  = \frac{2\ell+1}{4\pi} \frac{1}{4!}
 \left(   \rule{0pt}{14pt}
  v^{\{3} v^3 v^3 v^{3\}_0} P_\ell^{(4)}(x)
  + 6 \verp{\{3}\verpp{3} v^3 v^{3\}_0} P_\ell^{(3)}(x) \right. \\
  &+3 \verp{\{3} \verp{3} \verpp{3} \verpp{3\}_0} P_\ell^{(2)}(x) 
-\frac{12}{7} (6\ell(\ell+1)-5)
\left( v^{\{3} v^{3\}_0} P_\ell^{(2)}(x) +
\verp{\{3} \verpp{3\}_0} P_\ell^{(1)}(x) \right)\\
 &\left. + \frac{8}{5} \ell (\ell+1) (3\ell(\ell+1)-1)
 P_\ell(x) \right),
  \end{aligned}\\
  \label{eq:momrese}  
 & \begin{aligned}
\sum_{\ell_z=-\ell}^\ell &{\ell_z}^5 Y_{\ell \ell_z}(\verpp{})
  Y_{\ell \ell_z}(\verp{})^*  = i \frac{2\ell+1}{4\pi} \frac{1}{5!}
  \left(\rule{0pt}{14pt}  
  v^{\{3} v^3 v^3 v^3 v^{3\}_0} P_\ell^{(5)}(x)
  + 10\, \verp{\{3}\verpp{3} v^3 v^3 v^{3\}_0} P_\ell^{(4)}(x)
  \right. \\
  &+ 15\, \verp{\{3} \verp{3} \verpp{3} \verpp{3} v^{3\}_0}
  P_\ell^{(2)}(x)  - \frac{100}{9} (2\ell(\ell+1)-3) 
  \left( v^{\{3} v^3 v^{3\}_0} P_\ell^{(3)}(x)  + 3\, \verp{\{3} \verpp{3}
  v^{3\}_0} P_\ell^{(2)}(x) \right)\\
& \left.+\frac{120}{7} (3\ell^2(\ell+1)^2 - 3\ell (\ell+1) +1) v^3
P_\ell^{(1)}(x) \right).
  \end{aligned}
\end{align}
\end{subequations}
In the last two equalities the tensors on the r.h.s.\ may be
numerically evaluated as, e.g., $v^{\{3} v^3 v^3 v^{3\}_0} =
4!\veps{3333}(0)^* \veps{i_1i_2i_3i_4}(0) v^{i_1} v^{i_2} v^{i_3}
v^{i_4}$, as we actually have done for numerical verifications of
(\ref{eq:momresd}), (\ref{eq:momrese}).

A second class of bilocal sums is obtained from (\ref{eq:mix1}) with
$n=0$ by multiplying both sides by $\veps{i_1\ldots i_t}(\pm t)$ and
using (I.\ref{I.eq:tensor2f}) and the known matrix elements of $L^\pm
= L^1\pm i L^2$.  Thus,
\begin{equation}
  \label{eq:addthr2}\renewcommand{\minalignsep}{0pt}
  \begin{aligned}
  &\sum_{\ell_z=-\ell}^\ell Y_{\ell(\ell_z\pm t)}(\verpp{})
  Y_{\ell\ell_z}(\verp{})^* &
  &\sqrt{\frac{(\ell\mp\ell_z)!}{(\ell\mp\ell_z-t)!}} 
  \sqrt{\frac{(\ell\pm\ell_z+t)!}{(\ell\pm\ell_z)!}} =
  \\
  &
  \sum_{\ell_z=-\ell}^\ell Y_{\ell(\ell_z\pm t)}(\verpp{})
  Y_{\ell\ell_z}(\verp{})^* &
  &\CG{\ell}{\ell_z}{t}{\pm t}{\ell}{\ell_z\pm t}
  (\pm1)^t 2^{\frac{t}{2}}
  \langle \ell || \veps{i_1\ldots i_t}L^{i_1}\ldots L^{i_t} || \ell
  \rangle =\\
  & & &\hspace{22ex}
  =i^t \frac{2\ell+1}{4\pi} \sum_{q=0}^{[t/2]} C_{t,q} (\verp{\pm})^q
  (\verpp{\pm})^q (v^\pm)^{t-2q} P_\ell^{(t-q)}(x),
  \end{aligned}
\end{equation}
with $C_{t,q}$ as defined in (I.\ref{I.eq:genL}), $y^\pm = y^1\pm iy^2
= \mp\sqrt{2} \veps{}(\pm1)\cdot\vec{y}$ for any $\vec{y}$, and with
$Y_{\ell m}\equiv 0$ if $|m|>\ell$.  It is understood in
(\ref{eq:addthr2}) that we must choose either all upper signs or all
lower ones.

Next, we consider (\ref{eq:mix1}) with $t=0$, so only matrix elements
of $\verr{}$ are involved.  In that situation (\ref{eq:mix1})
generalizes to the bilocal case some results known in the local case
$\verpp{} = \verp{}$ (see \cite{var88} and refs.\ therein). We give
here their explicit form for $n=1$, 2.  For $n=1$, $t=0$ we have,
\begin{subequations}
  \label{eq:rmat6}
  \begin{align}
  \label{eq:rmat6a}
&\sum_{\ell_z=-\ell}^\ell Y_{\ell_1\ell_z}(\verpp{})
Y_{\ell\ell_z}(\verp{})^* \sqrt{\left(
    \lave + 1/2\right)^2 - \ell_z^2} =
\dl \frac{1}{4\pi} \sqrt{2\ell+1} \sqrt{2\ell_1+1} \left(
\verpp{3} P'_{\ell_1}(x) - \verp{3} P'_{\ell}(x)
  \right),
\\
  \label{eq:rmat6b}
&\begin{aligned}
\lefteqn{
\sum_{\ell_z=-\ell}^\ell Y_{\ell_1(\ell_z+m)}(\verpp{})
Y_{\ell\ell_z}(\verp{})^* \sqrt{\ell_1+\dl m \ell_z} 
\sqrt{\ell_1 + \dl m \ell_z + 1} =}
\hspace{45ex}\\
&=-\frac{m}{4\pi} \sqrt{2\ell+1} \sqrt{2\ell_1+1} \left(
  \verpp{m} P'_{\ell_1}(x) - \verp{m} P'_{\ell}(x)\right),
\end{aligned}
  \end{align}
\end{subequations}
where $m=\pm1$.  For $n=2$, $t=0$ we can write (\ref{eq:mix1}) as
\begin{subequations}
  \label{eq:rmat7}
  \begin{align}
    \label{eq:rmat7a}
&    \begin{aligned}
\sum_{\ell_z=-\ell+1}^{\ell-1} Y&_{(\ell+1)\ell_z}(\verpp{})
Y_{(\ell-1)\ell_z}(\verp{})^*
\sqrt{(\ell^2-\ell_z^2)((\ell+1)^2-\ell_z^2)} =
\hspace{15ex}\\
& =\frac{1}{4\pi} \sqrt{(2\ell-1)(2\ell+3)} \left(
(\verpp{3})^2 P''_{\ell+1}(x) -2 \verp{3} \verpp{3} P''_{\ell}(x)  
+ (\verp{3})^2 P''_{\ell-1}(x) - P'_\ell(x)\right),
    \end{aligned}\\
    \label{eq:rmat7b}
& \begin{aligned}
\sum_{\ell_z=-\ell+1}^{\ell-1}
Y&_{(\ell+1)(\ell_z+m)}(\verpp{}) Y_{(\ell-1)\ell_z}(\verp{})^*
\sqrt{(\ell^2-\ell_z^2)}\sqrt{(\ell+m\ell_z+1)
(\ell+m\ell_z+2)} = \\
& = -\frac{m}{4\pi} \sqrt{(2\ell-1)(2\ell+3)} \left(
  \verpp{m}\verpp{3} 
P''_{\ell+1}(x) - (\verpp{m} \verp{3} + \verpp{3} \verp{m})
P''_{\ell}(x) + \verp{m}\verp{3} P''_{\ell-1}(x) \right),
\end{aligned}\\
    \label{eq:rmat7c}
& \begin{aligned} 
\sum_{\ell_z=-\ell+1}^{\ell-1}
Y&_{(\ell+1)(\ell_z+2m)}(\verpp{}) Y_{(\ell-1)\ell_z}(\verp{})^*
\sqrt{\frac{(\ell+m\ell_z+3)!}{(\ell+m\ell_z-1)!}} = \\
& = \frac{1}{4\pi} \sqrt{(2\ell-1)(2\ell+3)} \left( (\verpp{m})^2 
 P''_{\ell+1}(x) - 2 \verpp{m}\verp{m}
 P''_{\ell}(x) + (\verp{m})^2 P''_{\ell-1}(x) \right),
    \end{aligned}
  \end{align}
\end{subequations}
where on the second and third equalities $m=\pm1$, and $\verp{m} =
\verp{1} + i m \verp{2}$.  For $n>2$, similar but more involved
results are obtained from (\ref{eq:mix1}).  We will not discuss them
further here for reasons of space.

Finally, we discuss some explicit forms of (\ref{eq:mix1}) for low
values of $n,t>0$, which extend (\ref{eq:rmat6}) and (\ref{eq:rmat7})
by insertion of powers of $\ell_z$ in the l.h.s.  For $n=1=t$ with
$m=0$, $\pm1$ we find
\begin{subequations}
\label{eq:mix3}
  \begin{align}
\label{eq:mix3a}
&\sum_{\ell_z=-\ell}^\ell Y_{\ell_1\ell_z}(\verpp{})
Y_{\ell\ell_z}(\verp{})^* \ell_z \sqrt{(\lave + 1/2)^2-\ell_z^2} 
= 
i \frac{\dl}{4\pi} \sqrt{(2\ell+1)(2\ell_1+1)}\, v^3 \left(
\verpp{3} P''_{\ell_1}(x) - \verp{3} P''_{\ell}(x)\right),
\\
\label{eq:mix3b}
&\begin{aligned}
\lefteqn{
\sum_{\ell_z=-\ell}^\ell Y_{\ell_1(\ell_z+m)}(\verpp{})
Y_{\ell\ell_z}(\verp{})^* \ell_z \sqrt{\frac{(\ell_1+\dl m
  l_z+1)!}{(\ell_1+\dl m l_z-1)!}}  
= -\frac{1}{8\pi} \sqrt{(2\ell+1)(2\ell_1+1)}
\left[ i m (v^m \verpp{3} +
\rule{0pt}{10pt}\right.}
\hspace{4.5ex}\\
& + v^3 \verpp{m}) \left.P''_{\ell_1}(x) -
i m (v^m \verp{3} + v^3 \verp{m}) P''_{\ell}(x)
+\dl (\ell+(1-\dl)/2) \left(
\verpp{m} P'_{\ell_1}(x) - \verp{m} P'_{\ell}(x)
\right)\rule{0pt}{10pt}\right],
\end{aligned}
  \end{align}
\end{subequations}
to be compared with (\ref{eq:rmat6}).  Similarly, for $n=1$, $t=2$,
with $m=0$ we get
\begin{equation}
  \label{eq:mix4}
  \begin{aligned}
\lefteqn{
\sum_{\ell_z=-\ell}^\ell Y_{\ell_1\ell_z}(\verpp{})
Y_{\ell\ell_z}(\verp{})^* {\ell_z}^2 \sqrt{(\lave + 1/2)^2-\ell_z^2} 
= -\frac{\dl}{20\pi} \sqrt{(2\ell+1)(2\ell_1+1)} \left[
\left(-\verpp{3} (1-x^2) + 5 \verpp{3} (v^3)^2\right)\right.
}\hspace{12ex}\\
&\times 
P'''_{\ell_1}(x) -
\left(-\verp{3} (1-x^2) + 5 \verp{3} (v^3)^2\right)
P'''_{\ell}(x) 
+\left(-\verp{3} - 2x \verpp{3} + 5 \verp{3} (\verpp{3})^2 \right) 
P''_{\ell_1}(x) \\
&\left.
-\left(-\verpp{3} - 2x \verp{3} + 5 \verpp{3} (\verp{3})^2 \right) 
P''_{\ell}(x) 
-(\lave-1/2)(\lave+3/2) \left(\verpp{3} P'_{\ell_1}(x) -
 \verp{3} P'_\ell(x)\right) \right],
  \end{aligned}
\end{equation}
to be compared with (\ref{eq:rmat6a}) and (\ref{eq:mix3a}).  The
analogous results for $m=\pm1$, $\pm2$, $\pm3$ are somewhat lengthy,
so we omit them for brevity.  Finally, with $n=2$, $t=1$ and $m=0$
we obtain
\begin{equation}
  \label{eq:mix5}
  \begin{aligned}
\lefteqn{
\sum_{\ell_z=-\ell+1}^{\ell-1} Y_{(\ell+1)\ell_z}(\verpp{})
Y_{(\ell-1)\ell_z}(\verp{})^* \ell_z \sqrt{(\ell^2-\ell_z^2)
((\ell+1)^2-\ell_z^2)}  
=  
\frac{1}{4\pi} \sqrt{(2\ell-1)(2\ell_1+3)}\, i v^3 }
\hspace{35ex}\\
&\times\left[
(\verpp{3})^2 P'''_{\ell+1}(x) - 2 \verpp{3} \verp{3} P'''_\ell(x) +
(\verp{3})^2 P'''_{\ell-1}(x) - P''_\ell(x)
\right],
  \end{aligned}
\end{equation}
which is related to (\ref{eq:rmat7a}).  As in the previous case we
omit the explicit results for $m\neq 0$, as well as those for higher
$n$ and $t$, for brevity.  Those cases can be obtained from the
general form (\ref{eq:mix2}), either algebraically or numerically.
Analogous, but lengthier, results can be obtained by the same
techniques for higher values of $n$ and $t$.  

Before leaving the subject of bilocal sums of spherical harmonics, we
briefly discuss the restriction of the previous results to the local
case $\verpp{} = \verp{}$.
From (\ref{eq:mix2}) with (I.\ref{I.eq:mix4}), (I.\ref{I.eq:tensor6}),
(I.\ref{I.eq:redmat7}) and (I.\ref{I.eq:redmat4}) we get,
\begin{equation}
  \label{eq:mix6}
  \begin{aligned}
\lefteqn{ \sum_{\ell_z=-\ell}^\ell Y_{\ell_n(\ell_z+m)}(\verp{})
  Y_{\ell \ell_z}(\verp{})^*
  \CG{\ell}{\ell_z}{n+t}{m}{\ell_n}{\ell_z+m}=
}\hspace{41ex}\\
&= \sqrt{\frac{(2\ell+1)(2\ell_n+1)}{2(n+t)+1}}
\CG{\ell}{0}{n+t}{0}{\ell_n}{0} \frac{1}{\sqrt{4\pi}}
Y_{(n+t)m}(\verp{}),
  \end{aligned}
\end{equation}
valid for $n,t\ge0$ and $|m| \leq n+t$.  Notice that the
r.h.s.\ vanishes for $t$ odd, because the CG coefficient does.  This
is due to the fact that, as seen from (I.\ref{I.eq:mix4}), when
$\verpp{}=\verp{}$ and $t$ is odd all terms in the r.h.s.\ of
(\ref{eq:mix1}), (\ref{eq:mix2}) contain a positive tensor power of
$\vec{v}=0$.  Equation (\ref{eq:mix6}) is, of course, the inverse of
the well-known Clebsch-Gordan series for spherical harmonics
\cite{var88}.  For particular values of its parameters it reduces to
sums of spherical harmonics with certain weight functions.  We
list several of those particular cases in appendix
\ref{sec:applocal}.

\section{Addition theorems for one scalar and one tensor spherical
  harmonic} 
\label{sec:scalspin}

The results from the previous section for bilocal sums of spherical
harmonics lead to a completely general addition theorem involving one
scalar and one tensor spherical harmonic.  Multiplying both sides of
(\ref{eq:mix2}) by $\veps{i_1\ldots i_nj_1\ldots j_t}(m)^*$, and
summing over $m$, we get
\begin{equation}
  \label{eq:sca-ten-addthr1}
  \begin{split}
  \sum_{\ell'_z=-\ell_n}^{\ell_n} Y_{\ell_n\ell'_z}(\verpp{})\left(
  \left.Y_{\ell_n\ell'_z}^{\ell(n+t)}(\verp{})\right)^{i_1\ldots
    i_{n+t}}\right.^* =
  (-1)^n \sqrt{\frac{2\ell_n+1}{2\ell+1}} \sum_{\ell_z=-\ell}^\ell 
  \left(Y_{\ell\ell_z}^{\ell_n (n+t)}(\verpp{})\right)^{i_1\ldots
    i_{n+t}} Y_{\ell\ell_z}(\verp{})^* \\
= \frac{1}{(n+t)!} 
\frac{\langle\verpp{}|\P_{\ell_n} \verr{\{i_1}\ldots \verr{i_n}
  L^{i_{n+1}} \ldots L^{i_{n+t}\}_0} \P_\ell | \verp{} \rangle}{
\langle\ell_n||\veps{k_1\ldots k_{n+t}} \verr{k_1}\ldots \verr{k_n}
  L^{k_{n+1}} \ldots L^{k_{n+t}} ||\ell\rangle}~,
  \end{split}
\end{equation}
with $n$, $t$ nonnegative integers.  The matrix elements on the
r.h.s.\ of (\ref{eq:sca-ten-addthr1}) are given explicitly by
(I.\ref{I.eq:mix4}) and (I.\ref{I.eq:redmat7}) for $n>0$, $t>0$, by
(I.\ref{I.eq:genL}) and (I.\ref{I.eq:redmat3}) for $n=0$, $t>0$, and
by (I.\ref{I.eq:rmat3}) and (I.\ref{I.eq:redmat6}) for $n>0$, $t=0$.
Notice that, in the last case, the matrix element in
(\ref{eq:sca-ten-addthr1}) involves an additional symmetrization with
respect to (I.\ref{I.eq:rmat3}).  We give here the expanded forms of
(\ref{eq:sca-ten-addthr1}) for spins $n+t=1$, 2, 3 (see \cite{var88}
for some related results).

For vector spherical harmonics (\ref{eq:sca-ten-addthr1}) yields,
\begin{subequations}
  \label{eq:sca-ten-addthr2}
  \begin{align}
    \label{eq:sca-ten-addthr2a}
\sum_{\ell_z=-\ell}^\ell \left(Y^{\ell 1}_{\ell\ell_z} (\verpp{})
\right)^i Y_{\ell\ell_z}(\verp{})^* &=
\frac{i}{4\pi} \frac{2\ell+1}{\sqrt{\ell(\ell+1)}}\, v^i P'_\ell(x)~,
\\ 
    \label{eq:sca-ten-addthr2b}
\sum_{\ell_z=-\ell}^\ell \left(Y^{\ell_1 1}_{\ell\ell_z} (\verpp{})
\right)^i Y_{\ell\ell_z}(\verp{})^* &=
-\frac{\sqrt{2}}{4\pi} \sqrt{\frac{2\ell+1}{\ell_1+\ell+1}} 
\left(\verpp{i} P'_{\ell_1}(x) - \verp{i} P'_\ell(x)\right).
  \end{align}
\end{subequations}
For rank-2 tensor spherical harmonics, by setting $n+t=2$ in
(\ref{eq:sca-ten-addthr1}) we get,
\begin{subequations}
\label{eq:sca-ten-addthr3}
\begin{align}
  \label{eq:sca-ten-addthr3a}
\sum_{\ell_z=-\ell}^\ell \left(Y^{\ell 2}_{\ell\ell_z} (\verpp{})
\right)^{i_1 i_2} Y_{\ell\ell_z}(\verp{})^* &=
-\frac{\sqrt{3/2}}{4\pi} \frac{(2\ell+1)/\sqrt{\ell(\ell+1)}}{
  \sqrt{(2\ell-1)(2\ell+3)}}
  \left( v^{\{i_1} v^{i_2\}_0} P''_\ell(x)
  + \verpp{\{i_1} \verp{i_2\}_0} P'_\ell(x)\right), \\   
  \label{eq:sca-ten-addthr3b}
\sum_{\ell_z=-\ell}^\ell \left(Y^{\ell_1 2}_{\ell\ell_z} (\verpp{})
\right)^{i_1 i_2} Y_{\ell\ell_z}(\verp{})^* &=
-\frac{i}{2\pi} \sqrt{2\ell+1} \sqrt{\frac{(\ell_1+\ell-3)!!}
  {(\ell_1+\ell+3)!!}}\left( \verpp{\{i_1} v^{i_2\}} P''_{\ell_1}(x) -
  \verp{\{i_1} v^{i_2\}} P''_\ell(x) \right), \\
  \label{eq:sca-ten-addthr3c}
\sum_{\ell_z=-\ell}^\ell \left(Y^{\ell_2 2}_{\ell\ell_z} (\verpp{})
\right)^{i_1 i_2} Y_{\ell\ell_z}(\verp{})^* &=
\frac{1}{4\pi}\frac{1}{\sqrt{2\ell_2+1}}
\sqrt{\frac{(\ell_2+\ell-1)(\ell_2+\ell+3)}
  {(\ell_2+\ell)(\ell_2+\ell+1)(\ell_2+\ell+2)}}\\
&\quad\times\nonumber
\left(
\verpp{\{i_1} \verpp{i_2\}_0} P''_{\ell_2}(x) - 2 \verpp{\{i_1} 
\verp{i_2\}_0} P''_{\ell_1}(x) + \verp{\{i_1} \verp{i_2\}_0}
P''_\ell(x) \right).
\end{align}
\end{subequations}
For rank-3 tensor spherical harmonics, setting $n+t=3$ in
(\ref{eq:sca-ten-addthr1}) yields,
\begin{subequations}
\label{eq:sca-ten-addthr4}
\begin{align}
  \label{eq:sca-ten-addthr4a}
&\begin{aligned}
\lefteqn{
\sum_{\ell_z=-\ell}^\ell \left(Y^{\ell 3}_{\ell\ell_z} (\verpp{})
\right)^{i_1 i_2 i_3} Y_{\ell\ell_z}(\verp{})^* =
-\frac{i}{2\pi} \frac{\sqrt{10}}{3} \sqrt{\frac{(2\ell-3)!}
  {(2\ell+4)!}} (2\ell+1)^{3/2}
}\hspace{40ex} \\
&\times 
\left(
v^{\{i_1} v^{i_2} v^{i_3\}_0} P^{(3)}_\ell(x) + 3
\verpp{\{i_1} \verp{i_2} v^{i_3\}_0} P^{(2)}_\ell(x)
\right),
\end{aligned}\\
  \label{eq:sca-ten-addthr4b}
&\begin{aligned}
\lefteqn{
\sum_{\ell_z=-\ell}^\ell \left(Y^{\ell_1 3}_{\ell\ell_z} (\verpp{}) 
\right)^{i_1 i_2 i_3} Y_{\ell\ell_z}(\verp{})^* =
\frac{1}{4\pi} \sqrt{\frac{10}{3}} \sqrt{2\ell+1}
\sqrt{\frac{(\ell_1+\ell-3)!}{(\ell_1+\ell+4)!}} \sqrt{(\ell_1+\ell)
  (\ell_1+\ell+2)} 
}\hspace{3ex}\\
&\times
\left(
  \verpp{\{i_1} v^{i_2} v^{i_3\}_0} P^{(3)}_{\ell_1}(x) 
  - \verp{\{i_1} v^{i_2} v^{i_3\}_0} P_\ell^{(3)}(x)
 + \verpp{\{i_1} \verpp{i_2} \verp{i_3\}_0}
  P_{\ell_1}^{(2)}(x) - \verpp{\{i_1} \verp{i_2} \verp{i_3\}_0}
  P_{\ell}^{(2)}(x) \right),
\end{aligned}\\
  \label{eq:sca-ten-addthr4c}
&\begin{aligned}
\lefteqn{
\sum_{\ell_z=-\ell}^\ell \left(Y^{\ell_2 3}_{\ell\ell_z} (\verpp{}) 
\right)^{i_1 i_2 i_3} Y_{\ell\ell_z}(\verp{})^* =
\frac{i}{8\pi} \frac{1}{\sqrt{3}}
\frac{1}{(2\ell_1+1)\sqrt{2\ell_2+1}} 
\sqrt{\frac{(2\ell_1+3)!!}{(2\ell_1-3)!!}}
\sqrt{\frac{(\ell_1-2)!}{(\ell_1+2)!}} 
}\hspace{25ex}\\
&\times
\left(
  \verpp{\{i_1} \verpp{i_2} v^{i_3\}_0} P^{(3)}_{\ell_2}(x) 
  - 2 \verpp{\{i_1} \verp{i_2} v^{i_3\}_0} P_{\ell_1}^{(3)}(x)
  + \verp{\{i_1} \verp{i_2} v^{i_3\}_0} P_{\ell}^{(3)}(x)
  \right),
\end{aligned}\\
  \label{eq:sca-ten-addthr4d}
&\begin{aligned}
\lefteqn{
\sum_{\ell_z=-\ell}^\ell \left(Y^{\ell_3 3}_{\ell\ell_z} (\verpp{})  
\right)^{i_1 i_2 i_3} Y_{\ell\ell_z}(\verp{})^* =
-\frac{1}{3\pi}
\frac{\sqrt{2}\sqrt{\lave+2}}{(2\ell_1+1)(2\ell_2+1)\sqrt{2\ell_3+1}}  
\sqrt{\frac{(\ell_3+\ell-3)!!}{(\ell_3+\ell+3)!!}} 
\sqrt{\frac{(\lave+1)!}{(\lave-2)!}}
}\hspace{0ex}  \\
&\times
\left(
\verpp{\{i_1} \verpp{i_2} \verpp{i_3\}_0} P^{(3)}_{\ell_3}(x) 
-3 \verpp{\{i_1} \verpp{i_2} \verp{i_3\}_0} P^{(3)}_{\ell_2}(x) 
+3 \verpp{\{i_1} \verp{i_2} \verp{i_3\}_0} P^{(3)}_{\ell_1}(x) 
- \verp{\{i_1} \verp{i_2} \verp{i_3\}_0} P^{(3)}_{\ell}(x) 
\right).
\end{aligned}
\end{align}
\end{subequations}
Similar relations for tensor spherical harmonics of arbitrary rank
$n+t>3$ are contained in (\ref{eq:sca-ten-addthr1}).

Equation (\ref{eq:sca-ten-addthr1}) yields also information on matrix
elements of operators between spinless and integer-spin states.  We
consider, as the simplest representative example, matrix elements of
the operator $L^3$.  Combining (\ref{eq:sca-ten-addthr1}), with
$n+t=1$, with the decomposition into irreducible components of the
tensor matrix elements $\langle \ell,\ell'_z | L^i L^j | \ell, \ell_z
\rangle$ and $\langle \ell,\ell'_z | \verr{i} L^j | \ell, \ell_z
\rangle$, and setting $j=3$, we get the following bilocal sums of one
scalar and one vector spherical harmonic weighted by $\ell_z$
\begin{subequations}
  \label{eq:sca-ten-addthr5}
  \begin{align}
  \label{eq:sca-ten-addthr5a}
&\begin{aligned}
\lefteqn{
\sum_{\ell_z=-\ell}^{\ell} \ell_z \left( Y_{\ell\ell_z}^{\ell 1}
(\verpp{}) \right)^i  Y_{\ell\ell_z}(\verp{})^* =
-\frac{1}{8\pi} \frac{2\ell+1}{\sqrt{\ell(\ell+1)}} 
\left( v^{\{i} v^{3\}_0} P^{(2)}_\ell(x) + \verpp{\{i} \verp{3\}_0}
  P'_\ell(x)\right.
}\hspace{53ex} \\
&\left. +(\verp{i} \verpp{3} - \verp{3} \verpp{i}) P'_\ell(x) 
- \frac{2}{3} \ell (\ell+1) \delta^{i3} P_\ell(x)
\right),
\end{aligned}\\
  \label{eq:sca-ten-addthr5b}
&\begin{aligned}
\lefteqn{
\sum_{\ell_z=-\ell}^{\ell} \ell_z \left( Y_{\ell\ell_z}^{\ell_1 1}
(\verpp{}) \right)^i  Y_{\ell\ell_z}(\verp{})^* =
  -\frac{i}{8\pi}\sqrt{2}\sqrt{\frac{2\ell+1}{\ell_1+\ell+1}}
  \left(\rule{0pt}{15pt} 
     \left(
      \verpp{\{i} v^{3\}} P''_{\ell_1}(x) - \verp{\{i}
    v^{3\}} P''_\ell(x) \right)\right.
}\hspace{32ex}\\
&  \left. -\left((\ell_1-\ell)(\ell+1/2)-1/2)\right)
\varepsilon^{i3p} \left(\verpp{p} P'_{\ell_1}(x) - \verp{p}
  P'_\ell(x)\right)   \rule{0pt}{15pt} \right).
  \end{aligned}
  \end{align}
\end{subequations}
Similarly, from (\ref{eq:sca-ten-addthr1}) and the irreducible-tensor
decompositions of $\langle \ell,\ell'_z | L^i L^j L^k | \ell, \ell_z
\rangle$ and $\langle \ell,\ell'_z | \verr{i} L^j L^k | \ell, \ell_z
\rangle$, setting $j=3=k$ we obtain bilocal sums of one scalar and one
vector spherical harmonic weighted by $\ell_z^2$
\begin{subequations}
\label{eq:sca-ten-addthr6}
\begin{align}
  \label{eq:sca-ten-addthr6a}
&\begin{aligned}
  \lefteqn{ \sum_{\ell_z=-\ell}^\ell \ell_z^2
\left(
Y_{\ell \ell_z}^{\ell 1}(\verpp{})\right)^i {Y_{\ell
\ell_z}(\verp{})}^* = -\frac{i}{4\pi}
\frac{2\ell+1}{\sqrt{\ell(\ell+1)}} 
\left(\frac{1}{6} \left( v^{\{i} v^3 v^{3\}_0} P_\ell^{(3)}(x) + 3
    \verpp{\{i} \verp{3} v^{3\}_0} P_\ell^{(2)}(x)\right)\right.
}\hspace{20ex}\\
&+ \left(\verp{i}\verpp{3} - \verp{3} \verpp{i} \right) v^3
P_\ell^{(2)}(x) + \frac{1}{2} \varepsilon^{i3h} \verpp{\{h}
\verp{3\}_0} P_\ell'(x) - \frac{1}{5} (\ell(\ell+1)+1/2) v^i
P'_\ell(x) \\
&- \left. \frac{2}{5} (\ell(\ell+1)-3/4) v^3 \delta^{i3}
P'_\ell(x) \rule{0pt}{15pt}\right), 
\end{aligned}\\
  \label{eq:sca-ten-addthr6b}
  &\begin{aligned} 
\lefteqn{ \sum_{\ell_z=-\ell}^\ell \ell_z^2 \left(
Y_{\ell \ell_z}^{\ell_11}(\verpp{})\right)^i {Y_{\ell
\ell_z}(\verp{})}^* = -\frac{1}{2\pi}
\sqrt{\frac{\ell+1/2}{\ell_1+\ell+1}} \left( -\frac{1}{6} \left(
\verpp{\{i} v^3 v^{3\}_0} P^{(3)}_{\ell_1}(x) -
\verp{\{i} v^3 v^{3\}_0} P^{(3)}_{\ell}(x) \right.\right.
}\hspace{6ex}\\
&+ \left. \verpp{\{i} \verpp{3} \verp{3\}_0} P''_{\ell_1}(x) -
 \verp{\{i} \verp{3} \verpp{3\}_0} P''_{\ell}(x)\right)
+\frac{1}{6} \left((\ell_1-\ell)(\ell_1+\ell+1)-3\right)
\varepsilon^{i3p}\\ 
&\times \left( \verpp{\{p} 
 v^{3\}_0} P''_{\ell_1}(x) - \verp{\{p} v^{3\}_0}
 P''_{\ell}(x) \right) 
- \frac{1}{20} (\ell_1(\ell_1+1)-9\ell(\ell+1)-2) \left(
\verpp{i} P'_{\ell_1}(x) - \verp{i} P'_{\ell}(x)
\right)\\
&+ \left.\frac{1}{20} (3\ell_1(\ell_1+1)-7\ell(\ell+1)-6) \left(
\verpp{3} P'_{\ell_1}(x) - \verp{3} P'_{\ell}(x) \right)
\delta^{i3}\right).
\end{aligned}
\end{align}
\end{subequations}
Finally, combining (\ref{eq:sca-ten-addthr1}) with the
irreducible-tensor decompositions of $\langle \ell,\ell'_z | L^i L^j
L^k | \ell, \ell_z \rangle$, $\langle \ell,\ell'_z | \verr{i} L^j L^k
| \ell, \ell_z \rangle$ and $\langle \ell,\ell'_z | \verr{i} \verr{j}
L^k | \ell, \ell_z \rangle$, with $k=3$, we obtain bilocal sums of one
scalar and one rank-2 tensor spherical harmonic weighted by $\ell_z$
\begin{subequations}
  \label{eq:sca-ten-addthr7}
  \begin{align}
    \label{eq:sca-ten-addthr7a}
&\begin{aligned}
  \lefteqn{
\sum_{\ell_z=-\ell}^\ell \ell_z
\left(
Y_{\ell \ell_z}^{\ell 2}(\verpp{})\right)^{ij} {Y_{\ell
\ell_z}(\verp{})}^* = 
-\frac{i}{2\pi} \sqrt{\frac{3}{2}} 
\frac{(2\ell+1)/\sqrt{\ell(\ell+1)}}{\sqrt{(2\ell-1)(2\ell+3)}}  
\left[\rule{0pt}{0pt}
\frac{1}{6} v^{\{i}v^{j}v^{3\}_0} P^{(3)}_\ell(x) 
\right.
+ \frac{1}{4} \left(\rule{0pt}{14pt}\!
-\varepsilon^{ijh} v^{\{h}v^{3\}_0}\right.
}\hspace{9ex}\\
& \left. + 2 \verpp{\{i} \verp{j} v^{3\}_0} +
  \frac{4}{3} \left(
    (\verp{i}\verpp{j}-\verp{j}\verpp{i}) v^3  
+ (\verp{j}\verpp{3}-\verp{3}\verpp{j}) v^i  
+ 2 (\verp{i}\verpp{3}-\verp{3}\verpp{i}) v^j \right)
\right) P^{(2)}_\ell(x)\\
&+\left(
\frac{1}{6}\left( -\frac{1}{2} \varepsilon^{ijh} 
\verpp{\{h}\verp{3\}_0} + \varepsilon^{hj3}  
\verpp{\{h}\verp{i\}_0} + 2 \varepsilon^{i3h} 
\verpp{\{h}\verp{j\}_0} \right)
-\frac{1}{4} \varepsilon^{ijh} (\verp{h}\verpp{3} - 
\verp{3}\verpp{h}) \right.\\
&\left.\left.+ \frac{2}{15} (\ell(\ell+1)-3/4) \delta^{ij} v^3 
-\frac{1}{5} (\ell(\ell+1)+1/2) \delta^{j3} v^i 
-\frac{1}{5} (\ell(\ell+1)-2) \delta^{i3} v^j 
\right) P'_\ell(x)
\rule{0pt}{0pt}\right],
\end{aligned}\\  
    \label{eq:sca-ten-addthr7b}
&\begin{aligned}
  \lefteqn{ \sum_{\ell_z=-\ell}^\ell \ell_z
\left(
Y_{\ell \ell_z}^{\ell_1 2}(\verpp{})\right)^{ij} {Y_{\ell
  \ell_z}(\verp{})}^* = \frac{\sqrt{2\ell+1}}{\pi}
\sqrt{\frac{(\ell_1+\ell-3)!!}{(\ell_1+\ell+3)!!}}
\left[\rule{0pt}{14pt}
\frac{1}{6} \left( \verpp{\{i} v^j v^{3\}_0} P^{(3)}_{\ell_1}(x) 
- \verp{\{i} v^j v^{3\}_0} P^{(3)}_{\ell}(x) 
\right.\right.}\hspace{8ex}\\  
&+ \left. \verpp{\{i} \verpp{j} \verp{3\}_0} P^{(2)}_{\ell_1}(x) 
- \verpp{\{i} \verp{j} \verp{3\}_0} P^{(2)}_{\ell}(x) \right)
+\frac{1}{24} (\ell_1(\ell_1+1)-\ell(\ell+1)-6)
\left(
(\varepsilon^{ih3} \verpp{\{h} v^{j\}_0} \right.\\
&+ \left. \varepsilon^{jh3}
\verpp{\{h} v^{i\}_0} ) P^{(2)}_{\ell_1}(x) -
(\varepsilon^{ih3} \verp{\{h} v^{j\}_0} + \varepsilon^{jh3}
\verp{\{h} v^{i\}_0} ) P^{(2)}_{\ell}(x)\right)
-\frac{3}{80}(\ell_1+\ell-1)(\ell_1+\ell+3) \\
&\times  
\left.\left( 
(\delta^{j3} \verpp{i} + \delta^{i3} \verpp{j} 
-2/3\, \delta^{ij} \verpp{3}) P'_{\ell_1}(x) 
-(\delta^{j3} \verp{i} + \delta^{i3} \verp{j} 
-2/3\,\delta^{ij} \verp{3}) P'_{\ell}(x) \right)
\rule{0pt}{14pt}\right],
\end{aligned}\\
    \label{eq:sca-ten-addthr7c}
&\begin{aligned}
\lefteqn{
  \sum_{\ell_z=-\ell}^\ell \ell_z \left(
Y_{\ell\ell_z}^{\ell_22}(\verpp{})\right)^{ij}
{Y_{\ell\ell_z}(\verp{})}^* = \frac{i}{24\pi} 
\sqrt{\frac{2\ell+1}{2\ell_1+1}}
\frac{1}{\sqrt{\ell_1(\ell_1+1)}} 
\left[\rule{0pt}{11pt}
\verpp{\{i} \verpp{j} v^{3\}_0}
P^{(3)}_{\ell_2}(x) 
- 2\verpp{\{i} \verp{j} v^{3\}_0}
P^{(3)}_{\ell_1}(x)\right.}
\hspace{31ex}\\ 
&+
\verp{\{i} \verp{j} v^{3\}_0} P^{(3)}_{\ell}(x) 
- \frac{1}{2} \left((\ell_1-\ell) (\ell_2 + \ell +1) -3\right) 
(\varepsilon^{hj3}\delta^{pi} - \varepsilon^{ih3} \delta^{pj}) \\
&\left.\times
\left(
\verpp{\{h} \verpp{p\}_0} P^{(2)}_{\ell_2}(x) - 
2 \verpp{\{h} \verp{p\}_0} P^{(2)}_{\ell_1}(x) +
\verp{\{h} \verp{p\}_0} P^{(2)}_{\ell}(x)
\right)\right]~.
  \end{aligned}
  \end{align}
\end{subequations}
Similar results involving higher powers of $\ell_z$ and/or higher-rank
tensor spherical harmonics follow by considering higher-rank
operators.  Also with the same techniques, other components of
$\vec{L}$, instead of $L^3$, can be considered, or other operators.

\subsection{Local results}
\label{sec:scalspinlocal}

The previous results simplify considerably when specialized to
$\verpp{}=\verp{}$.  The local case is of interest by itself, so we
discuss it in this section.   The general addition theorem
(\ref{eq:sca-ten-addthr1}) with $\verpp{}=\verp{}$ reduces to
\begin{equation}
  \label{eq:sca-ten-local0}
  \begin{split}
  \sum_{\ell'_z=-\ell_n}^{\ell_n} Y_{\ell_n \ell'_z}(\verp{}) 
  {\left(Y_{\ell_n\ell'_z}^{\ell(n+t)}(\verp{})\right)^{i_1\ldots
  i_nj_1\ldots j_t}}^* = (-1)^n \sqrt{\frac{2\ell_n+1}{2\ell+1}}
  \sum_{\ell_z=-\ell}^\ell
  \left(Y_{\ell\ell_z}^{\ell_n(n+t)}(\verp{})\right)^{i_1\ldots 
  i_nj_1\ldots j_t} {Y_{\ell_n \ell_z}(\verp{})}^* \\
  =(-1)^{n+t} \sqrt{(2\ell+1)(2\ell_n+1)}
  \CG{\ell}{0}{n+t}{0}{\ell_n}{0} \frac{1}{\sqrt{4\pi}} \left(
  Y^{(n+t)(n+t)}_{0 0}(\verp{})\right)^{i_1\ldots 
  i_nj_1\ldots j_t},
  \end{split}
\end{equation}
a result valid for all $\ell$, $n$, $t\ge 0$.  Notice that the r.h.s.\
vanishes for $t$ odd, for the same reasons as explained in connection
with (\ref{eq:mix6}).  Thus, for vector spherical harmonics, $n+t=1$
in (\ref{eq:sca-ten-local0}), the only non-trivial result is 
\begin{equation}
  \label{eq:sca-ten-local1}
\sum_{\ell_z=-\ell}^\ell \left(Y^{\ell_1 1}_{\ell
    \ell_z}(\verp{})\right)^i {Y_{\ell\ell_z}(\verp{})}^* =
-\frac{(\ell_1-\ell)}{4\pi}\sqrt{(\ell+1/2)(\ell_1+\ell+1)}
\,\verp{i}~. 
\end{equation}
For rank-2 tensor spherical harmonics, the local version of
(\ref{eq:sca-ten-addthr3}) is
\begin{subequations}
  \label{eq:sca-ten-local2}
\begin{align}
  \label{eq:sca-ten-local2a}
\sum_{\ell_z=-\ell}^\ell \left(Y^{\ell 2}_{\ell
    \ell_z}(\verp{})\right)^{j_1j_2} {Y_{\ell\ell_z}(\verp{})}^*
&=
-\frac{1}{8\pi} \sqrt{\frac{3}{2}} (2\ell+1) \sqrt{\frac{\ell(\ell+1)}
{(2\ell-1)(2\ell+3)}} \,\verp{\{j_1} \verp{j_2\}_0}~,\\
  \label{eq:sca-ten-local2b}
\sum_{\ell_z=-\ell}^\ell \left(Y^{\ell_2 2}_{\ell
    \ell_z}(\verp{})\right)^{i_1i_2} {Y_{\ell\ell_z}(\verp{})}^*
&=
\frac{3}{16\pi}  \sqrt{\ell_1(\ell_1+1)}
\sqrt{\frac{2\ell+1}{\ell_2+\ell+1}} \,\verp{\{i_1} \verp{i_2\}_0}~,
\end{align}
\end{subequations}
with (\ref{eq:sca-ten-addthr3b}) trivially vanishing for
$\verpp{}=\verp{}$.   For rank-3 tensor spherical harmonics, the
non-trivial local versions of (\ref{eq:sca-ten-addthr4}) are
\begin{subequations}
  \label{eq:sca-ten-local3}
\begin{align}
  \label{eq:sca-ten-local3a}
&\begin{aligned}
\sum_{\ell_z=-\ell}^\ell \left(Y^{\ell_1 3}_{\ell
    \ell_z}(\verp{})\right)^{i_1j_1j_2} {Y_{\ell\ell_z}(\verp{})}^*
&=
\frac{\ell_1-\ell}{64\pi} \sqrt{\frac{10}{3}} \sqrt{2\ell+1} 
\sqrt{\frac{(\ell_1+\ell-1)(\ell_1+\ell+1)(\ell_1+\ell+3)}
{(\ell_1+\ell-2)(\ell_1+\ell+4)}} \\
&\quad \times \verp{\{i_1} \verp{j_1} \verp{j_2\}_0}~,
\end{aligned}\\
  \label{eq:sca-ten-local3b}
&\begin{aligned}
\sum_{\ell_z=-\ell}^\ell \left(Y^{\ell_3 3}_{\ell
    \ell_z}(\verp{})\right)^{i_1i_2i_3} {Y_{\ell\ell_z}(\verp{})}^* 
&=
-\frac{\ell_1-\ell}{64\pi} \frac{5\sqrt{2}}{3} \sqrt{2\ell+1} 
\sqrt{\frac{(\ell_3+\ell-1)(\ell_3+\ell+1)(\ell_3+\ell+3)}
{(\ell_3+\ell)(\ell_3+\ell+2)}}\\
&\quad \times \verp{\{i_1} \verp{i_2} \verp{i_3\}_0}~.
\end{aligned}
\end{align}
\end{subequations}
The matrix elements of $L^3$ between one scalar and one vector
spherical harmonic, with $\verpp{}=\verp{}$, are given by the
local form of (\ref{eq:sca-ten-addthr5}) 
\begin{subequations}
  \label{eq:sca-ten-local4}
  \begin{align}
  \label{eq:sca-ten-local4a}
\sum_{\ell_z=-\ell}^\ell \ell_z \left(Y^{\ell 1}_{\ell
    \ell_z}(\verp{})\right)^i {Y_{\ell\ell_z}(\verp{})}^*
&= -\frac{1}{8\pi}(2\ell+1)\sqrt{\ell(\ell+1)} \left(
\verp{i} \verp{3} - \delta^{i3}
\right),\\
  \label{eq:sca-ten-local4b}
\sum_{\ell_z=-\ell}^\ell \ell_z \left(Y^{\ell_1 1}_{\ell
    \ell_z}(\verp{})\right)^i {Y_{\ell\ell_z}(\verp{})}^*
&= \frac{i}{8\pi}\sqrt{2}\sqrt{\frac{2\ell+1}{\ell_1+\ell+1}} \ell
  (\ell+1)  \varepsilon^{i3h} \verp{h}.
\end{align}
\end{subequations}
For $(L^3)^2$, the local analog of (\ref{eq:sca-ten-addthr6}) is
\begin{subequations}
  \label{eq:sca-ten-local5}
  \begin{align}
  \label{eq:sca-ten-local5a}
&\begin{aligned}
\sum_{\ell_z=-\ell}^\ell {\ell_z}^2 \left(Y^{\ell 1}_{\ell
    \ell_z}(\verp{})\right)^i {Y_{\ell\ell_z}(\verp{})}^*
&= -\frac{i}{8\pi} (2\ell+1)\sqrt{\ell(\ell+1)}\,
\varepsilon^{i3h}\verp{h} \verp{3},
\end{aligned}\\
  \label{eq:sca-ten-local5b}
&\begin{aligned}
\lefteqn{
\sum_{\ell_z=-\ell}^\ell {\ell_z}^2 \left(Y^{\ell_1 1}_{\ell
    \ell_z}(\verp{})\right)^i {Y_{\ell\ell_z}(\verp{})}^*
= \frac{(\ell_1-\ell)}{2\pi}\sqrt{(\ell+1/2)(\ell_1+\ell+1)} 
\left[
\frac{1}{96} (\ell_1+\ell-1)(\ell_1+\ell+3) \verp{\{i} \verp{3}
\verp{3\}_0} \right.
}\hspace{22ex}\\
&+\left.\frac{1}{40} (\ell_1(\ell_1+1)-9 \ell(\ell+1) -2) \verp{i} 
-\frac{1}{40} (3\ell_1(\ell_1+1)-7 \ell(\ell+1) -6) \verp{3} 
\delta^{i3}
\right].
\end{aligned}
\end{align}
\end{subequations}
Finally, for rank-2 tensor spherical harmonics, the local version of 
(\ref{eq:sca-ten-addthr7}) yields
\begin{subequations}
  \label{eq:sca-ten-local6}
  \begin{align}
  \label{eq:sca-ten-local6a}
&
\begin{aligned}
\lefteqn{
\sum_{\ell_z=-\ell}^\ell \ell_z \left(Y^{\ell 2}_{\ell
    \ell_z}(\verp{})\right)^{ij} {Y_{\ell\ell_z}(\verp{})}^* 
= -\frac{i}{24\pi}\sqrt{\frac{3}{2}}
(2\ell+1)\sqrt{\frac{\ell(\ell+1)}
{(2\ell-1)(2\ell+3)}} \left[
-\frac{1}{2} \varepsilon^{ijh} \verp{\{h} \verp{3\}_0}
+ \varepsilon^{hj3} \verp{\{h} \verp{i\}_0}
\right.
}\hspace{81ex}\\
&+ \left.
  2\varepsilon^{i3h} \verp{\{h} \verp{j\}_0} 
\rule{0pt}{14pt}\right],
\end{aligned}\\
  \label{eq:sca-ten-local6b}
&\begin{aligned}
\lefteqn{
\sum_{\ell_z=-\ell}^\ell \ell_z \left(Y^{\ell_1 2}_{\ell
    \ell_z}(\verp{})\right)^{ij} {Y_{\ell\ell_z}(\verp{})}^*
=
\frac{\ell_1-\ell}{32\pi} \sqrt{2\ell+1} 
\sqrt{(\ell_1+\ell-1)(\ell_1+\ell+1)(\ell_1+\ell+3)}
\left[\rule{0pt}{14pt}
\frac{1}{3}\verp{\{i} \verp{j} \verp{3\}_0}
\right.
}\hspace{65ex} \\
&- \left. \frac{3}{5} \left(
\delta^{j3} \verp{i} + \delta^{i3} \verp{j} 
- \frac{2}{3} \delta^{ij} \verp{3}\right)
\right],
\end{aligned}\\
  \label{eq:sca-ten-local6c}
&\begin{aligned}
\lefteqn{
\sum_{\ell_z=-\ell}^\ell \ell_z \left(Y^{\ell_2 2}_{\ell
    \ell_z}(\verp{})\right)^{ij} {Y_{\ell\ell_z}(\verp{})}^*
=
-\frac{i}{32\pi} ((\ell_1-\ell)(2\ell+1)-1) 
\sqrt{\frac{2\ell+1}{2\ell_1+1}} 
\sqrt{\ell_1(\ell_1+1)} \left( \varepsilon^{i3h} \verp{\{h}
  \verp{j\}_0} \right.
}\hspace{83ex} \\
&+ \left. \varepsilon^{j3h} \verp{\{h}
  \verp{i\}_0} \right).
\end{aligned}
\end{align}
\end{subequations}

\section{Addition theorems for spin spherical harmonics}
\label{sec:spin}

We consider now addition theorems for spin spherical harmonics of the
form 
\begin{equation}
  \label{eq:seed}
  \sum_{j_z=-j}^j Y_{jj_z}^{\ell' s'}(\verrp{}) {Y_{jj_z}^{\ell
      s}(\verr{})}^*, 
\end{equation}
with $s',s>0$.  Combining the factorization of
orbital and spin degrees of freedom in products of two CG coefficients
from sect.\ I.\ref{I.sec:fac} and the matrix elements of orbital
operators from sect.\ I.\ref{I.sec:proj}, we have
\begin{equation}
  \label{eq:adthrgen}
  \begin{aligned}
\lefteqn{
\sum_{j_z=-j}^j Y^{\ell's'}_{jj_z}(\verpp{}) 
{Y^{\ell s}_{jj_z}(\verp{})}^*  =
\sum_{\Delta=\Delta_\mathrm{min}}^{\Delta_\mathrm{max}} 
\frac{1}{\Delta!} C^{s's\Delta}_{\ell'\ell j}
\langle\verpp{}| \P_{\ell'} \verr{\{i_1} \ldots
\verr{i_{|\Delta\ell|}} L^{i_{|\Delta\ell|+1}} \ldots L^{i_\Delta\}_0} 
\P_\ell | \verp{} \rangle
}\hspace{59ex}\\
&\times \langle\psi_{s'} | T^{i_1} \ldots
T^{i_{|\Delta\ell|}} S^{i_{|\Delta\ell|+1}} \ldots S^{i_\Delta} 
 | \psi_s \rangle.
 \end{aligned}
\end{equation}
In this equality $\P_\ell$ stands for the orbital angular momentum
projector operator (see sect.\ I.\ref{I.sec:proj}), $|\psi_s\rangle$
are the basis spin states entering the definition of $Y^{\ell s}_{j
  j_z}$ (see sect.\ \ref{sec:ssh}), and $C^{s's\Delta}_{\ell'\ell j}$
and $\Delta_{\substack{\mathrm{max}\\\mathrm{min}}}$ are defined in 
(I.\ref{I.eq:CGmaster}).  The orbital matrix element appearing in 
(\ref{eq:adthrgen}) is given explicitly for any values of their
parameters in sect.\ I.\ref{I.sec:proj}.  For low values of $s',s>0$,
the spin matrix elements and simplified forms of the coefficients 
$C^{s's\Delta}_{\ell'\ell j}$ are given in sect.\ I.\ref{I.sec:fac}.
In the remainder of this section we discuss the resulting addition
theorems for spin spherical harmonics.  As above, throughout this
section  we denote $x\equiv \verp{}\cdot\verpp{}$ and 
$\vec{v}\equiv \verp{}\wedge \verpp{}$. 

\subsection{Addition theorems for spin-1/2 spherical harmonics}
\label{sec:spin1half}

Spin-1/2 spherical harmonics $Y_{jj_z}^{\ell\frac{1}{2}}$, as defined
by (\ref{eq:3halfdef}), vanish identically unless $j=\ell\pm 1/2$.  In
this subsection we always assume those values for $j$.  In the case of
two spin-1/2 harmonics with the same orbital angular momentum,
(\ref{eq:adthrgen}) together with (\ref{eq:mix1}) with $n=0$ yield
\begin{equation}
  \label{eq:add1half2}
  \begin{aligned}
  \sum_{j_z=-j}^j \left(Y_{jj_z}^{\ell\frac{1}{2}}(\verpp{})\right)_A
  {\left(Y_{jj_z}^{\ell\frac{1}{2}}(\verp{})\right)_B}^* &=
  \frac{j+1/2}{2\ell+1} \langle\verpp{}|\P_\ell|\verp{}\rangle
  \delta_{AB} + \frac{2(j-\ell)}{2\ell+1}
  \langle\verpp{}|L^k \P_\ell|\verp{}\rangle \sigma^k_{AB}~,\\
  &= \frac{j+1/2}{4\pi} P_\ell(x) \delta_{AB} + i
  \frac{2(j-\ell)}{4\pi} P'_\ell(x)\, \vec{v} \cdot
  \vec{\sigma}_{AB}~,
  \end{aligned}
\end{equation}
where in the second equality we used the matrix elements
(I.\ref{I.eq:proj3}) and (I.\ref{I.eq:L}). Eq. (\ref{eq:add1half2}) is
the addition theorem for spin-1/2 spherical harmonics with the same
orbital angular momentum.  It plays an important role in the
partial-wave expansion of (parity conserving) $S$-matrix elements
between spin-1/2 states with the same intrinsic parity (see, e.g.,
\cite{nac90}).  If the intrinsic parity of the initial and final
spin-1/2 states is opposite, on the other hand, a parity-conserving
transition can occur only if the initial and final orbital angular
momenta differ by one unit. To derive the addition theorem for
spin-1/2 spherical harmonics with different orbital angular momenta we
substitute (I.\ref{I.eq:1half-3}) for the product of CG coefficients,
(\ref{eq:mix1}) with $t=0$, and (I.\ref{I.eq:rmat4a}) for the
resulting matrix element of $\P_\ell$, in (\ref{eq:adthrgen}) to get
\begin{equation}
  \label{eq:add1half3}
  \sum_{j_z=-j}^j \left(Y_{jj_z}^{\ell_1\frac{1}{2}}(\verpp{})\right)_A
  {\left(Y_{jj_z}^{\ell\frac{1}{2}}(\verp{})\right)_B}^* =
  -\frac{\ell_1-\ell}{4\pi} \left( \verpp{}\cdot \vec{\sigma}_{AB}
    P'_{\ell_1} (x) - \verp{}\cdot \vec{\sigma}_{AB}
    P'_\ell(x) \right).
\end{equation}
This equation holds for either $\ell_1=\ell+1$ and $j=\ell+1/2$, or
$\ell_1=\ell-1$ and $j=\ell-1/2$, otherwise the l.h.s.\ trivially
vanishes.   The local forms of these addition theorems are also of
interest.  Setting $\verpp{}=\verp{}$ in the above equalities we get,
\begin{equation}
  \label{eq:add1halflocal}
\sum_{j_z=-j}^j \left(Y_{jj_z}^{\ell\frac{1}{2}}(\verp{})\right)_A
  {\left(Y_{jj_z}^{\ell\frac{1}{2}}(\verp{})\right)_B}^* =  
\frac{2j+1}{8\pi}  \delta_{AB},
\quad
\sum_{j_z=-j}^j \left(Y_{jj_z}^{\ell_1\frac{1}{2}}(\verp{})\right)_A 
  {\left(Y_{jj_z}^{\ell\frac{1}{2}}(\verp{})\right)_B}^* =   
-\frac{2\lave+1}{8\pi}  \verp{}\cdot\vec{\sigma}_{AB}~.
\end{equation}
Eqs.\ (\ref{eq:add1half2}) and (\ref{eq:add1half3}) exhaust
the addition theorems of the form (\ref{eq:seed}) for two spin-1/2
spherical harmonics.

\subsection{Addition theorems for spin-1 spherical harmonics}
\label{sec:spin1}

We consider first the addition theorem for two spin-1 spherical
harmonics with the same orbital angular momentum.  We proceed in the
same way as in the previous section, with the separation of orbital
and spin quantum numbers in the product of CG coefficients carried out
by means of (I.\ref{I.eq:1-2}) and the orbital matrix elements given
by (\ref{eq:mix1}) with $n=0$, (I.\ref{I.eq:L}) and (I.\ref{I.eq:LL}),
to find, 
\begin{equation}
  \label{eq:add1-1}
  \begin{aligned}
\lefteqn{
  \sum_{j_z=-j}^j \left(Y_{jj_z}^{\ell 1}(\verpp{})\right)^r
  {\left(Y_{jj_z}^{\ell 1}(\verp{})\right)^s}^* =
}\hspace{12ex} \\
&= \frac{2\ell+1}{4\pi} \left[
C^{110}_{\ell\ell j} P_\ell(x) \delta^{rs} 
+ C^{111}_{\ell\ell j} P'_\ell(x) \verp{[r}\verpp{s]}
+ \frac{1}{2} C^{112}_{\ell\ell j} 
\left(
P'_\ell(x) \verp{\{r}\verpp{s\}_0} + P''_\ell(x) v^{\{r} v^{s\}_0}
\right)\right],
  \end{aligned}
\end{equation}
with $C^{11p}_{\ell\ell j}$, $p=0,1,2$, as given by
(I.\ref{I.eq:1-2}).  Eq.\ (\ref{eq:add1-1}) obviously can hold only
when $j=\ell$, $\ell\pm 1$, its l.h.s.\ trivially vanishing otherwise.

For vector spherical harmonics with orbital angular momenta differing
by one unit, with the product of CG coefficients in (I.\ref{I.eq:1-5})
and the matrix elements (\ref{eq:mix1}) with $n=1$, $t=0$ and $n=1=t$,
and (I.\ref{I.eq:rmat4a}), (I.\ref{I.eq:mix5a}), we find
\begin{equation}
  \label{eq:add1-2}  
  \begin{aligned}
\lefteqn{
  \sum_{j_z=-j}^j \left(Y_{jj_z}^{\ell_1 1}(\verpp{})\right)^r
  {\left(Y_{jj_z}^{\ell 1}(\verp{})\right)^s}^* = 
}\hspace{8ex}\\
&= -\frac{i}{8\pi}(\ell_1-\ell) \left[ 2 C^{111}_{\ell_1\ell j}
    \varepsilon^{rsh} \left(\verpp{h} P'_{\ell_1} (x) - \verp{h}
      P'_{\ell} (x) \right) + C^{112}_{\ell_1\ell j}
   \left(v^{\{r} \verpp{s\}} P''_{\ell_1} (x) - v^{\{r} \verp{s\}}
     P''_{\ell} (x) \right)\right],
  \end{aligned}
\end{equation}
with the coefficients $C^{11p}_{\ell_1\ell j}$, $p=1,2$, defined in
(I.\ref{I.eq:1-5}). We remark here that (\ref{eq:add1-2}) holds if
either $\ell_1=\ell+1$ and $j=\ell+1$, $\ell$, or $\ell_1=\ell-1$ and
$j=\ell$, $\ell-1$, otherwise the l.h.s.\ vanishes identically.

Similarly, for vector spherical harmonics with orbital angular momenta
differing by two units, with the product of CG coefficients in eq.\
(I.\ref{I.eq:1-8}) and the matrix elements (\ref{eq:mix1}) with
$n=2$, $t=0$ and (I.\ref{I.eq:rmat4b}) we get
\begin{equation}
  \label{eq:add1-3}
  \sum_{j_z=-j}^j \left(Y_{jj_z}^{\ell_2 1}(\verpp{})\right)^r
  {\left(Y_{jj_z}^{\ell 1}(\verp{})\right)^s}^* = -\frac{1}{4\pi}
  \frac{1}{\sqrt{j(j+1)}}
\left(
    \verpp{(r}\verpp{s)_0} P''_{\ell_2}(x)
    -\verpp{\{r}\verp{s\}_0} P''_{\ell_1}(x)
    +\verp{(r}\verp{s)_0} P''_{\ell}(x)
    \right).
\end{equation}
In this equation $x$ is defined as in (\ref{eq:add1-1}). Equation
(\ref{eq:add1-3}) holds only if either $\ell_2=\ell+2$ and
$j=\ell_1=\ell+1$, or $\ell_2=\ell-2$ and $j=\ell_1=\ell-1$, the
l.h.s.\ being identically zero in any other case.   

In the local case $\verpp{}=\verp{}$ the above results take the form
\begin{equation}
  \label{eq:add1-local}
  \begin{aligned}
\sum_{j_z=-j}^j \left(Y_{jj_z}^{\ell 1}(\verp{})\right)^r
{\left(Y_{jj_z}^{\ell 1}(\verp{})\right)^s}^* &= 
\frac{2\ell+1}{4\pi} \left[
C^{110}_{\ell\ell j}\delta^{rs} + \frac{1}{4} \ell(\ell+1) 
C^{112}_{\ell\ell j} \verp{\{r} \verp{s\}_0}
\right],\\
\sum_{j_z=-j}^j \left(Y_{jj_z}^{\ell_1 1}(\verp{})\right)^r
{\left(Y_{jj_z}^{\ell 1}(\verp{})\right)^s}^* &= 
\frac{i}{8\pi} \sqrt{2j+1}\sqrt{2j-\lave+1/2}\,
\varepsilon^{rsh} \verp{h}, 
\\
\sum_{j_z=-j}^j \left(Y_{jj_z}^{\ell_2 1}(\verp{})\right)^r
{\left(Y_{jj_z}^{\ell 1}(\verp{})\right)^s}^* &= 
-\frac{3}{8\pi} \sqrt{j(j+1)}\verp{(r}\verp{s)_0}. 
  \end{aligned}
\end{equation}
Eqs.\ (\ref{eq:add1-1}), (\ref{eq:add1-2}), (\ref{eq:add1-3}) are all
possible addition theorems of the form (\ref{eq:seed}) for two vector
spherical harmonics (see \cite{var88} for related results).

\subsection{Addition theorems for spin-3/2 spherical harmonics}
\label{sec:spin3half}

For spin-3/2 spherical harmonics we have four possible addition
theorems, with $0\leq\ell'-\ell \leq 3$.  We begin with the case
$\ell'=\ell$, for which the product of CG coefficients is given by
(I.\ref{I.eq:3h-2}), and the needed orbital matrix elements are
(\ref{eq:mix1}) with $n=0$ and (I.\ref{I.eq:L})---(I.\ref{I.eq:LLL}).
With those results we arrive at,
\begin{equation}
  \label{eq:add3hal1}
  \begin{split}
\sum_{j_z=-j}^j \left(Y_{jj_z}^{\ell \frac{3}{2}}(\verpp{})\right)^i_A
  {\left(Y_{jj_z}^{\ell \frac{3}{2}}(\verp{})\right)^j_B}^* =
  \frac{2\ell+1}{4\pi} X^{ih}_{AC} \left[
    C^{\frac{3}{2}\frac{3}{2}0}_{\ell\ell j} P_\ell(x) \delta^{hk}
    \delta_{CD} 
+ i \frac{3}{2} C^{\frac{3}{2}\frac{3}{2}1}_{\ell\ell j} P'_\ell(x)
\delta^{hk} \vec{v}\cdot \vec{\sigma}_{CD}
+\frac{3}{2} C^{\frac{3}{2}\frac{3}{2}2}_{\ell\ell j} 
  \right.\\
\times\left. 
\left(v^{\{h} v^{k\}_0} P''_\ell(x) 
+ 
    \verp{\{h} \verpp{k\}_0} P'_\ell(x) \right) \delta_{CD}
  + \frac{i}{4} C^{\frac{3}{2}\frac{3}{2}3}_{\ell\ell j}
  \left(v^{\{h}v^k v^{t\}_0} P'''_\ell(x) + 3 \verp{\{h} \verpp{k}
    v^{t\}_0} P''_\ell(x) \right) \sigma^t_{CD} \right] X^{kj}_{DB},
  \end{split}
\end{equation}
where the coefficients $C^{\frac{3}{2}\frac{3}{2}p}_{\ell\ell j}$,
$p=0,\ldots, 3$ are given in (I.\ref{I.eq:3h-2b}), and $X^{pq}_{FG}$
is the projector defined in (I.\ref{I.eq:nspincompl}).  Notice that
when the r.h.s.\ of (\ref{eq:add3hal1}) is contracted with
$\vchiU{A}{i*}$ or $\vchiU{B}{j}$, the corresponding projector may be
dropped since it leaves the spinors invariant.  We remark that
(\ref{eq:add3hal1}) holds when $j=\ell\pm 3/2$, $\ell\pm 1/2$, its
l.h.s.\ vanishing identically otherwise.

For orbital angular momenta differing by one unit, the appropriate
factorization of orbital and spin degrees of freedom in CG
coefficients is (I.\ref{I.eq:3h-4}), and the orbital matrix elements 
are given by (I.\ref{I.eq:mix1}) with $t=0$ and (I.\ref{I.eq:rmat2}), 
(I.\ref{I.eq:mix1}) with $t=1$ and (I.\ref{I.eq:mix5a}), and
(I.\ref{I.eq:mix1}) with $t=2$ and (I.\ref{I.eq:mix5b}).  We then get, 
\begin{equation}
  \label{eq:add3hal2}
  \begin{aligned}
\lefteqn{
  \sum_{j_z=-j}^j \left(Y_{jj_z}^{\ell_1 \frac{3}{2}}(\verpp{})\right)^i_A
  {\left(Y_{jj_z}^{\ell \frac{3}{2}}(\verp{})\right)^j_B}^* =
  \frac{\ell_1-\ell}{4\pi} X^{ih}_{AC} \left[
    \frac{3}{2}C^{\frac{3}{2}\frac{3}{2}1}_{\ell_1\ell j}
 \left(\verpp{q} P'_{\ell_1}(x) - \verp{q}
    P'_\ell(x)\right) \delta^{hk} \sigma^q_{CD}
  \right.
}\hspace{2ex}\\  
  &- i \frac{3}{2} C^{\frac{3}{2}\frac{3}{2}2}_{\ell_1\ell j}
  \left(\verpp{\{h} v^{k\}_0}
    P''_{\ell_1}(x) - \verp{\{h} v^{k\}} P''_\ell(x)\right)
  \delta_{CD}
  + \frac{1}{4}C^{\frac{3}{2}\frac{3}{2}3}_{\ell_1\ell j} 
  \left( \verpp{\{h} v^k v^{q\}_0}
    P'''_{\ell_1}(x) - \verp{\{h} v^k v^{q\}_0} P'''_{\ell}(x)
  \right.\\
  &+ \left.\left. \verpp{\{h} \verpp{k} \verp{q\}_0}
      P''_{\ell_1}(x) - \verp{\{h} \verp{k} \verpp{q\}_0}
      P''_{\ell}(x) \right)\sigma^q_{CD} \right] X^{kj}_{DB}~,
  \end{aligned}
\end{equation}
with the coefficients $C^{\frac{3}{2}\frac{3}{2}p}_{\ell_1\ell j}$,
$p=1,\ldots, 3$ from (I.\ref{I.eq:3h-4b}) and the
projector $X^{pq}_{FG}$ defined in (I.\ref{I.eq:nspincompl})
Equation (\ref{eq:add3hal2}) is valid when either $\ell_1=\ell+1$ and
$j=\ell+3/2$, $\ell+1/2$, $\ell-1/2$, or $\ell_1=\ell-1$ and
$j=\ell+1/2$, $\ell-1/2$, $\ell-3/2$, its l.h.s.\ being identically
zero otherwise.

The case $|\ell'-\ell| = 2$ follows by using the relation among CG
coefficients  (I.\ref{I.eq:3h-6}), (\ref{eq:mix1}) with $n=2$,
$t=0$ and $t=1$, and the matrix elements (I.\ref{I.eq:rmat4b}) and
(I.\ref{I.eq:mix5b}), with 
the result
\begin{equation}
  \label{eq:add3hal3}
  \begin{aligned}
\lefteqn{
  \sum_{j_z=-j}^j \left(Y_{jj_z}^{\ell_2 \frac{3}{2}}(\verpp{})\right)^i_A
  {\left(Y_{jj_z}^{\ell \frac{3}{2}}(\verp{})\right)^j_B}^* =
  -\frac{1}{4\pi}\frac{1}{2\ell_1+1} X^{ip}_{AC}}
  \hspace{16ex}\\
  &\times\left[
    \frac{3}{2}C^{\frac{3}{2}\frac{3}{2}2}_{\ell_2\ell j}
    \left(\verpp{\{p} \verpp{r\}_0} 
      P''_{\ell_2}(x) + \verp{\{p} \verp{r\}_0} P''_{\ell}(x) 
      -2 \verpp{\{p} \verp{r\}_0} P''_{\ell_1}(x) \right) \delta_{CD}
  \right.\\
  &+ \left. \frac{i}{4} C^{\frac{3}{2}\frac{3}{2}3}_{\ell_2\ell j}
    \left( \verpp{\{p} \verpp{q} 
    v^{r\}_0} P'''_{\ell_2}(x) 
    +\verp{\{p} \verp{q} v^{r\}_0} P'''_{\ell}(x) -
    2\verpp{\{p} \verp{q} v^{r\}_0} P'''_{\ell_1}(x) \right)
    \sigma^{q}_{CD}\right] X^{rj}_{DB}~,
  \end{aligned}
\end{equation}
where $x$, $\vec{v}$ and $X^{pq}_{FG}$ are as above, and
$C^{\frac{3}{2}\frac{3}{2}p}_{\ell_2\ell j}$, $p=2$, 3, are defined in
(I.\ref{I.eq:3h-6}). Equation (\ref{eq:add3hal3}) is valid when either
$\ell_2=\ell+2$ and $j=\ell+3/2$, $\ell+1/2$, or $\ell_2=\ell-2$ and
$j=\ell-1/2$, $\ell-3/2$, its l.h.s.\ being identically zero
otherwise.

Finally, the case $|\ell'-\ell|=3$ follows from the CG relation
(I.\ref{I.eq:3h-8}) and (I.\ref{I.eq:mix1}) evaluated at $n=3$, $t=0$,
with (I.\ref{I.eq:rmat4c}).  With the same notations as in the
previous equations, we find
\begin{equation}
\label{eq:add3hal4}
\begin{aligned}
\lefteqn{
  \sum_{j_z=-j}^j \left(Y_{jj_z}^{\ell_3 \frac{3}{2}}(\verpp{})\right)^i_A
  {\left(Y_{jj_z}^{\ell \frac{3}{2}}(\verp{})\right)^j_B}^* =
  \frac{1}{4\pi}\frac{\ell_1-\ell}{(2\ell_1+1)(2\ell_2+1)}
  \sqrt{\frac{j(j+1)}{(j-1/2)(j+3/2)}} X^{ip}_{AC}}
  \hspace{0ex}\\
  &\times \left[
   \frac{1}{3} \verpp{\{p} \verpp{q} \verpp{r\}_0}
   P'''_{\ell_3}(x)
   -  \verpp{\{p} \verpp{q} \verp{r\}_0} P'''_{\ell_2}(x) 
   + \verpp{\{p} \verp{q} \verp{r\}_0} P'''_{\ell_1}(x)
   -\frac{1}{3} \verp{\{p} \verp{q} \verp{r\}_0} P'''_{\ell}(x) 
\right] \sigma^{q}_{CD} X^{rj}_{DB},
\end{aligned}
\end{equation}
which is valid for either $\ell_3 = \ell+3$ and $j=\ell+3/2$, or 
$\ell_3 = \ell-3$ and $j=\ell-3/2$.  For other values of $j$, the
l.h.s.\  vanishes.  

We quote here also the local forms of the previous results.  With
$\verpp{}=\verp{}$ we have
\begin{equation}
  \label{eq:add3hal-local}
  \begin{aligned}
\sum_{j_z=-j}^j \left(Y_{jj_z}^{\ell \frac{3}{2}}(\verp{})\right)^i_A
  {\left(Y_{jj_z}^{\ell \frac{3}{2}}(\verp{})\right)^j_B}^* &=
\frac{2\ell+1}{4\pi} X^{ih}_{AC} \left[
C^{\frac{3}{2}\frac{3}{2}0}_{\ell\ell j} \delta^{hk} + \frac{3}{4}
\ell(\ell+1) C^{\frac{3}{2}\frac{3}{2}2}_{\ell\ell j}
\verp{\{h} \verp{k\}_0}
\right] X^{kj}_{CB}~,
\\
\sum_{j_z=-j}^j \left(Y_{jj_z}^{\ell_1
    \frac{3}{2}}(\verp{})\right)^i_A {\left(Y_{jj_z}^{\ell
      \frac{3}{2}}(\verp{})\right)^j_B}^* &= \frac{1}{16\pi}
(\ell_1+\ell+1) X^{ih}_{AC} \left[ \rule{0pt}{14pt} 3
  C^{\frac{3}{2}\frac{3}{2}1}_{\ell_1\ell j} \verp{q} \delta^{hk}
\right.\\
&\quad \left.+\frac{1}{16} C^{\frac{3}{2}\frac{3}{2}3}_{\ell_1\ell j}
  (\ell_1+\ell-1)(\ell_1+\ell+3) \verp{\{h} \verp{k} \verp{q\}_0}
\right]  \sigma^{q}_{CD} X^{kj}_{DB}~,
\\
\sum_{j_z=-j}^j \left(Y_{jj_z}^{\ell_2 \frac{3}{2}}(\verp{})\right)^i_A
{\left(Y_{jj_z}^{\ell \frac{3}{2}}(\verp{})\right)^j_B}^* &=
-\frac{9}{16\pi} \frac{\ell_1(\ell_1+1)}{2\ell_1+1} 
C^{\frac{3}{2}\frac{3}{2}2}_{\ell_2\ell j} X^{ip}_{AC}
\verp{\{p} \verp{r\}_0} X^{rj}_{CB}~,
\\
\sum_{j_z=-j}^j \left(Y_{jj_z}^{\ell_3 \frac{3}{2}}(\verp{})\right)^i_A
{\left(Y_{jj_z}^{\ell \frac{3}{2}}(\verp{})\right)^j_B}^* &=
\frac{5}{192\pi} \frac{\sqrt{2j+1}}{(2\ell_1+1)(2\ell_2+1)} 
\sqrt{\frac{(2j+3)!}{(2j-2)!}} X^{ip}_{AC} \verp{\{p} \verp{q}
\verp{r\}_0} \sigma^{q}_{CD} X^{rj}_{DB}.
  \end{aligned}
\end{equation}
With (\ref{eq:add3hal1})--(\ref{eq:add3hal4})
established, there are no further non-trivial relations of the form
(\ref{eq:seed}) for two spin-3/2 spherical harmonics.

\subsection{Addition theorems for one spin-0 and one spin-1 spherical
  harmonics} 
\label{sec:add01}

Using the factorization of orbital and spin degrees of freedom in
products of CG coefficients provided by 
(I.\ref{I.eq:10-2})---(I.\ref{I.eq:10-5}), we
recover the results already given in (\ref{eq:sca-ten-addthr2}). 


\subsection{Addition theorems for one spin-1/2 and one spin-3/2 spherical
  harmonics} 
\label{sec:add13half}

For $s=1/2$ and $s'=3/2$, with $\ell'=\ell$, the relation among 
CG coefficients (I.\ref{I.eq:3h1h-2}) or (I.\ref{I.eq:3h1h-5}), and
(\ref{eq:mix1}) with $n=0$ and $t=1$, 2, together with
(I.\ref{I.eq:L}) and (I.\ref{I.eq:LL}), lead to, 
\begin{equation}
  \label{eq:add13half1}
  \begin{aligned}
\lefteqn{
  \sum_{j_z=-j}^j \left(Y_{jj_z}^{\ell \frac{3}{2}}(\verpp{})\right)^i_A
  {\left(Y_{jj_z}^{\ell \frac{1}{2}}(\verp{})\right)_B}^* =
  \frac{2\ell+1}{8\pi} \sqrt{\frac{3}{2}}X^{ik}_{AC} \left[
    C^{\frac{3}{2}\frac{1}{2}1}_{\ell\ell j} i v^k P'_\ell(x)
    \delta_{CB}\right. 
}\hspace{49ex}\\ 
   &-\left. \frac{1}{4} C^{\frac{3}{2}\frac{1}{2}2}_{\ell\ell j}
     \left(v^{\{k} v^{j\}_0} P''_\ell(x) + \verp{\{k} 
      \verpp{j\}_0}P'_{\ell}(x)\right) \sigma^j_{CB}
    \right],
  \end{aligned}
\end{equation}
with $C^{\frac{3}{2}\frac{1}{2}1,2}_{\ell_2\ell j}$ given by
(I.\ref{I.eq:3h1h-2b}) and $X^{pq}_{FG}$ by (I.\ref{I.eq:nspincompl}).
Notice that the subindex ``0'' in the second term on the r.h.s.\ of
(\ref{eq:add13half1}) is actually not needed, because $X^{ik}_{AC}
\sigma^k_{CB}\equiv 0$.  The case with $s=3/2$ and $s'=1/2$ follows
immediately from (\ref{eq:add13half1}) by complex conjugation and the
exchange $\verp{}\leftrightarrow\verpp{}$.  The equality in
(\ref{eq:add13half1}) holds for $j=\ell\pm 1/2$, otherwise
$Y_{jj_z}^{\ell \frac{1}{2}}(\verp{}) \equiv 0$.

The case $|\ell'-\ell|=1$ follows from (I.\ref{I.eq:3h1h-7}), 
(\ref{eq:mix1}) with $n=1$, $t=0$ and (I.\ref{I.eq:rmat4a}), and 
(\ref{eq:mix1}) with $n=1$, $t=1$ and (I.\ref{I.eq:mix5a})
\begin{equation}
  \label{eq:add13half2}
  \begin{aligned}
\lefteqn{
  \sum_{j_z=-j}^j \left(Y_{jj_z}^{\ell_1 \frac{3}{2}}(\verpp{})\right)^i_A
  {\left(Y_{jj_z}^{\ell \frac{1}{2}}(\verp{})\right)_B}^* =  
  \frac{\ell_1-\ell}{8\pi} \sqrt{\frac{3}{2}} X^{ik}_{AC} \left[
    C^{\frac{3}{2}\frac{1}{2}1}_{\ell_1\ell j} \left(\verpp{k}
      P'_{\ell_1}(x) - \verp{k} P'_\ell(x)\right) \delta_{CB}\right. 
}\hspace{50ex}\\
   &+\left. \frac{i}{4} C^{\frac{3}{2}\frac{1}{2}2}_{\ell_1\ell j} 
     \left(\verpp{\{k} v^{p\}} P''_{\ell_1}(x) - \verp{\{k} 
      v^{p\}}P''_{\ell}(x)\right) \sigma^p_{CB} \right],
  \end{aligned}
\end{equation}
with the coefficients $C^{\frac{3}{2}\frac{1}{2}1,2}_{\ell_1\ell j}$
given by (I.\ref{I.eq:3h1h-7b}), with $\Delta s=1$.  Eq.\
(\ref{eq:add13half2}) holds for 
$j=\ell\pm 1/2$, its l.h.s.\ vanishing otherwise.  Similarly,
\begin{equation}
  \label{eq:add13half2x}
  \begin{aligned}
\lefteqn{
\sum_{j_z=-j}^j \left(Y_{jj_z}^{\ell_1
\frac{1}{2}}(\verpp{})\right)_A {\left(Y_{jj_z}^{\ell
\frac{3}{2}}(\verp{})\right)^i_B}^* =
\frac{\ell_1-\ell}{8\pi} \sqrt{\frac{3}{2}} \left[
C^{\frac{1}{2}\frac{3}{2}1}_{\ell_1\ell j} \left(\verpp{k}
P'_{\ell_1}(x) - \verp{k}
P'_\ell(x)\right) \delta_{AC}\right.
}\hspace{43ex} \\
&+\left. \frac{i}{4}
C^{\frac{1}{2}\frac{3}{2}2}_{\ell_1\ell j}
 \left(\verpp{\{k} v^{p\}} P''_{\ell_1}(x) - \verp{\{k}
v^{p\}}P''_{\ell}(x)\right) \sigma^p_{AC}\right] X^{ki}_{CB}~,
  \end{aligned}
\end{equation}
where now $C^{\frac{1}{2}\frac{3}{2}1,2}_{\ell_1\ell j}$ are given by
(I.\ref{I.eq:3h1h-7b}) with $\Delta s=-1$, and the equality holds for
$j=\ell_1\pm 1/2$.

Finally, the last case to be considered is $\ell'-\ell=\pm2$.  In this
case we use the factorization of orbital and spin degrees of freedom
from (I.\ref{I.eq:3h1h-c}), together with (\ref{eq:mix1}) with $n=2$,
$t=1$ and (I.\ref{I.eq:rmat4b}).  In this way we get,
\begin{equation}
  \label{eq:add13half3}
  \begin{aligned}
\lefteqn{
  \sum_{j_z=-j}^j \left(Y_{jj_z}^{\ell_2 \frac{3}{2}}(\verpp{})\right)^i_A
  {\left(Y_{jj_z}^{\ell \frac{1}{2}}(\verp{})\right)_B}^* =  
  \frac{(\ell_1-\ell)}{4\pi} 
  \sqrt{\frac{j+1/2}{\ell_1(\ell_1+1)(2\ell_1+1)}} X^{ik}_{AC}
}\hspace{40ex} \\
  &\times \left[ \verpp{(k} \verpp{p)_0} P''_{\ell_2}(x) - \verpp{\{k}
      \verp{p\}_0} P''_{\ell_1}(x) + \verp{(k} \verp{p)_0}
      P''_{\ell}(x)\right]\sigma^p_{CB} ~,
  \end{aligned}
\end{equation}
with $X^{ik}_{AC}$ defined as in the previous equations.  The
result (\ref{eq:add13half3}) applies if either $\ell_2 = \ell+2$ and
$j=\ell+1/2$, or $\ell_2 = \ell-2$ and $j=\ell-1/2$, otherwise its
l.h.s.\ trivially vanishes.  Similarly, we can derive,
\begin{equation}
  \label{eq:add13half4}
  \begin{aligned}
\lefteqn{
  \sum_{j_z=-j}^j \left(Y_{jj_z}^{\ell_2 \frac{1}{2}}(\verpp{})\right)_A
  {\left(Y_{jj_z}^{\ell \frac{3}{2}}(\verp{})\right)^i_B}^* =  
  -\frac{(\ell_1-\ell)}{4\pi} 
  \sqrt{\frac{j+1/2}{\ell_1(\ell_1+1)(2\ell_1+1)}} \sigma^k_{AC}
}\hspace{40ex} \\
  &\times \left[ \verpp{(k} \verpp{p)_0} P''_{\ell_2}(x) - \verpp{\{k}
      \verp{p\}_0} P''_{\ell_1}(x) + \verp{(k} \verp{p)_0}
      P''_{\ell}(x)\right] X^{pi}_{CB} ~,
  \end{aligned}
\end{equation}
the extra minus sign in the r.h.s., with respect to
(\ref{eq:add13half3}), coming from the factor $\Delta s$ in 
(I.\ref{I.eq:3h1h-cb}). Now (\ref{eq:add13half4}) holds if either $\ell_2 = \ell+2$ and
$j=\ell+3/2$, or $\ell_2 = \ell-2$ and $j=\ell-3/2$.  

Setting $\verpp{}=\verp{}$ in the above equalities, we obtain the
local sums
\begin{equation}
  \label{eq:add13half-local}
  \begin{aligned}
\sum_{j_z=-j}^j \left(Y_{jj_z}^{\ell \frac{3}{2}}(\verpp{})\right)^i_A
{\left(Y_{jj_z}^{\ell \frac{1}{2}}(\verp{})\right)_B}^* &=
-\sqrt{\frac{3}{2}} \frac{2\ell+1}{64\pi} \ell(\ell+1) 
C^{\frac{3}{2}\frac{1}{2}2}_{\ell\ell j} X^{ik}_{AC}
\verp{\{k}\verp{j\}_0} \sigma^j_{CB}, \\
\sum_{j_z=-j}^j \left(Y_{jj_z}^{\ell_1 \frac{3}{2}}(\verpp{})\right)^i_A
{\left(Y_{jj_z}^{\ell \frac{1}{2}}(\verp{})\right)_B}^* &=  
\frac{1}{16\pi} \sqrt{\frac{3}{2}} (\ell_1+\ell+1) 
C^{\frac{3}{2}\frac{1}{2}1}_{\ell_1\ell j} X^{ik}_{AB} \verp{k},\\
\sum_{j_z=-j}^j \left(Y_{jj_z}^{\ell_1
\frac{1}{2}}(\verpp{})\right)_A {\left(Y_{jj_z}^{\ell
\frac{3}{2}}(\verp{})\right)^i_B}^* &=
\frac{1}{16\pi} \sqrt{\frac{3}{2}} (\ell_1+\ell+1) 
C^{\frac{1}{2}\frac{3}{2}1}_{\ell_1\ell j} \verp{k} X^{ki}_{AB},\\
\sum_{j_z=-j}^j \left(Y_{jj_z}^{\ell_2 \frac{3}{2}}(\verpp{})\right)^i_A
{\left(Y_{jj_z}^{\ell \frac{1}{2}}(\verp{})\right)_B}^* &=  
\frac{3}{16\pi}
(\ell_1-\ell)\sqrt{\frac{(j+1/2)\ell_1(\ell_1+1)}{2\ell_1+1}} 
X^{ik}_{AC} \verp{\{k} \verp{p\}_0} \sigma^p_{CB},\\
\sum_{j_z=-j}^j \left(Y_{jj_z}^{\ell_2 \frac{1}{2}}(\verpp{})\right)_A
{\left(Y_{jj_z}^{\ell \frac{3}{2}}(\verp{})\right)^i_B}^* &=  
-\frac{3}{16\pi}
(\ell_1-\ell)\sqrt{\frac{(j+1/2)\ell_1(\ell_1+1)}{2\ell_1+1}} 
\sigma^p_{AC}  \verp{\{p} \verp{k\}_0} X^{ki}_{CB}.
  \end{aligned}
\end{equation}
No further non-trivial addition theorems of the form (\ref{eq:seed})
can be given for one spin-1/2 and one spin-3/2 spherical harmonics.

\section{Final Remarks}
\label{sec:finrem}

Using the results of I we developed here a systematic approach to the
derivation of addition theorems for spin spherical harmonics.  The
results of our approach are appropriately summarized by the general
expressions (\ref{eq:mix1}), (\ref{eq:mix2}) for bilocal sums of
spherical harmonics, which are also explicit expressions for bilocal
spherical harmonics, the general form (\ref{eq:sca-ten-addthr1}) of
the addition theorem for one scalar and one tensor spherical
harmonics, and the general form (\ref{eq:adthrgen}) of the addition
theorem for two spin spherical harmonics.  Those results are based on
the irreducible-tensor matrix elements from sect.\ I.\ref{I.sec:proj}.
Bilocal sums of a more general type can be obtained by means of
reducible matrix elements, as illustrated by
(\ref{eq:momres})---(\ref{eq:mix5}) and
(\ref{eq:sca-ten-addthr5})---(\ref{eq:sca-ten-addthr7}).

In sect.\ \ref{sec:2scalar} we give results for bilocal sums of
ordinary spherical harmonics with certain weight functions depending
on $\ell_z$.  Many other classes of sums can in principle be obtained
with the same techniques, by considering other matrix elements.  In
sect.\ \ref{sec:scalspin} we explicitly state addition theorems for
and other bilocal sums of one scalar and one tensor spherical
harmonics, up to rank 3.  Results for higher ranks can equally well be
obtained, although the involved amount of algebraic labor is a rapidly
increasing function of rank.  The local versions of those results are
also of interest, and are summarized in sect.\
\ref{sec:scalspinlocal}.  Addition theorems for spin spherical
harmonics of the form (\ref{eq:seed}) are given in sect.\
\ref{sec:spin}, together with their local forms, for spins $1/2 \leq
s'=s \leq 3/2$, and $(s',s)=(3/2,1/2)$ and (1/2,3/2).  Similar results
for arbitrary values of $s',s$ can be obtained as particular cases of
(\ref{eq:adthrgen}) with the coefficients given by
(I.\ref{I.eq:CGmaster}) and the matrix elements from sect.\
I.\ref{I.sec:proj} and app.\ I.\ref{I.sec:matele}.  As in the
previous case, however, for higher spins ($s',s>2$) the algebraic
complexity of the results quickly grows unmanageable.  Clearly, for
spins larger than 2 an automated procedure would be needed.  Even
though there is no pressing physical need for such further
generalizations, because higher spin states occur infrequently in
nature, they could lead to improved calculational tools that would be
of interest even in phenomenological contexts.

The addition theorems for spin spherical harmonics given in sections
\ref{sec:scalspin} and \ref{sec:spin} are directly applicable to the
computation of partial-wave expansions of $S$-matrix elements for
two-body scattering of particles with spin.  As discussed in the
introduction that is our main motivation for this paper, but we
hope our results to be more widely applicable.

\appendix

\renewcommand{\theequation}{\thesection.\arabic{equation}}
\setcounter{equation}{0}

\section{Local sums of two spherical harmonics}
\label{sec:applocal}

In this appendix we list several particular cases of (\ref{eq:mix6}),
which are local versions of the bilocal sums of sect.\
\ref{sec:2scalar}.  Some of these relations are familiar from
textbooks and most, if not all, of them are certainly well known.  We
put them together here for reference (see also \cite{var88}).

For $n=0=t$, (\ref{eq:mix6}) reduces to the familiar equality
$\sum_{m=-\ell}^\ell \left|Y_{\ell m}(\theta,\varphi)\right|^2 =
(2\ell+1)/(4\pi)$ .  By setting $n=0=m$ in (\ref{eq:mix6}) with $t>0$
even, and proceeding as discussed in relation to (\ref{eq:momres}) we
obtain
\begin{subequations}
  \label{eq:momres2}
  \begin{align}
\sum_{\ell_z=-\ell}^\ell \left|Y_{\ell \ell_z}(\verr{})\right|^2 {\ell_z}^{2}
&=
\frac{1}{8\pi} (2\ell+1)\ell(\ell+1) (\sin\theta)^2~,
\label{eq:momres2b}\\ 
\sum_{{\ell_z}=-\ell}^\ell \left|Y_{\ell {\ell_z}}(\verr{})\right|^2 {\ell_z}^{4}
&= 
\frac{1}{8\pi} (2\ell+1)\ell(\ell+1) (\sin\theta)^2
\left( \frac{3}{4} (\ell(\ell+1)-2) (\sin\theta)^2 +1\right),
  \label{eq:momres2c}\\ 
\sum_{{\ell_z}=-\ell}^\ell \left|Y_{\ell {\ell_z}}(\verr{})\right|^2 {\ell_z}^{6}
&= 
\frac{1}{2^9\pi}(2\ell+1)\ell(\ell+1) (\sin\theta)^2
\left( \frac{5}{\ell(\ell+1)} \frac{(\ell+3)!}{(\ell-3)!}
  \cos(4\theta)
  \right.\nonumber\\
  &\quad\left.  - 20 \frac{(\ell+2)!}{(\ell-2)!} \cos(2\theta) + 15
    \ell^2 (\ell+1)^2 +4 \right),   \label{eq:momres2d}\\
  \sum_{{\ell_z}=-\ell}^\ell \left|Y_{\ell {\ell_z}}(\verr{})\right|^2 {\ell_z}^{8}
  &= \frac{35}{2^{14}\pi} (2\ell+1)(\sin\theta)^2 \left( -
    \frac{(\ell+4)!}{(\ell-4)!} \cos(6\theta) + 2
    \frac{(\ell+3)!}{(\ell-3)!} (3\ell(\ell+1)-4) \cos(4\theta)\right.
  \nonumber\\
  &\quad -\frac{1}{5} \frac{(\ell+2)!}{(\ell-2)!} (75 \ell^2
  (\ell+1)^2 - 70
  \ell(\ell+1) +24) \cos(2\theta)
  \nonumber\\
  &\quad \left. +\frac{2}{35} \ell(\ell+1) (175 \ell^3(\ell+1)^3 - 140
    \ell^2(\ell+1)^2 + 84 \ell (\ell+1) +16) \rule{0pt}{14pt}\right).
  \label{eq:momres2e}
  \end{align}
\end{subequations}
Higher moments of $\ell_z$ ($t>8$) can be computed, but the labor
involved increases rapidly with $t$.  The equalities
(\ref{eq:momres2b}) and (\ref{eq:momres2c}) are just the
particularization to $\verrp{}=\verr{}$ of (\ref{eq:momresb}) and
(\ref{eq:momresd}), resp.  When $m=\pm t$, with $t>0$ even,
(\ref{eq:mix6}) acquires the explicit form
\begin{equation}
\label{eq:rp=r.3}
\begin{split}
\lefteqn{\sum_{\ell_z=-\ell}^\ell Y_{\ell(\ell_z\pm t)}(\verr{})
Y_{\ell\ell_z}(\verr{})^* 
\sqrt{\frac{(\ell\mp\ell_z)!}{(\ell\mp\ell_z-t)!}}  
\sqrt{\frac{(\ell\pm\ell_z+t)!}{(\ell\pm\ell_z)!}}=}
\hspace{30ex}\\
&=\frac{(-1)^{t/2}}{4\pi} \frac{(t-1)!!}{2^{t/2}(t/2)!}
(2\ell+1) \frac{(\ell+t/2)!}{(\ell-t/2)!} (\sin\theta)^t
e^{\pm it\varphi}.   
\end{split}
\end{equation}
For odd $t$ the l.h.s.\ of this equation vanishes, in agreement with
(\ref{eq:mix6}).

For $t=0$, $n=1$ (\ref{eq:mix6}) reduces to,
\begin{subequations}
  \label{eq:rmat9}
  \begin{align}
  \label{eq:rmat9a}
\sum_{\ell_z=-\ell}^\ell Y_{(\ell+1)\ell_z}(\verp{})
Y_{\ell\ell_z}(\verp{})^* 
\sqrt{(\ell+1)^2 - \ell_z^2} &=
\frac{1}{4\pi} \sqrt{2\ell+1} \sqrt{2\ell+3}
\,(\ell+1) \cos\theta~,
\\
  \label{eq:rmat9b}
\sum_{\ell_z=-\ell}^\ell Y_{(\ell+1)(\ell_z+m)}(\verp{})
Y_{\ell\ell_z}(\verp{})^* \sqrt{\ell+1+ m \ell_z} 
\sqrt{\ell+2+ m \ell_z} &=
-\frac{m}{4\pi} \sqrt{2\ell+1} \sqrt{2\ell+3} 
\,(\ell+1) \sin\theta\, e^{im\varphi},
  \end{align}
\end{subequations}
where on the second line $m=\pm1$.  We restricted ourselves to
$\ell'=\ell+1$ in (\ref{eq:rmat9}), since the case $\ell'=\ell-1$ can
be obtained easily from there by appropriate transformation of
parameters.  Similarly, for $t=0$, $n=2$ we have
\begin{subequations}
  \label{eq:rmata}
  \begin{align}
    \label{eq:rmataa}
&    \begin{aligned}
\lefteqn{
\sum_{\ell_z=-\ell+1}^{\ell-1} Y_{(\ell+1)\ell_z}(\verp{})
Y_{(\ell-1)\ell_z}(\verp{})^*
\sqrt{(\ell^2-\ell_z^2)((\ell+1)^2-\ell_z^2)} =}
\hspace{45ex}\\
& =\frac{1}{8\pi} \sqrt{(2\ell-1)(2\ell+3)} \ell (\ell+1)
(3(\cos\theta)^2 -1),
    \end{aligned}\\
    \label{eq:rmatab}
& \begin{aligned}
\lefteqn{
\sum_{\ell_z=-\ell+1}^{\ell-1}
Y_{(\ell+1)(\ell_z+m)}(\verp{}) Y_{(\ell-1)\ell_z}(\verp{})^*
\sqrt{(\ell^2-\ell_z^2)}\sqrt{(\ell+m\ell_z+1)
(\ell+m\ell_z+2)} =}
\hspace{45ex} \\
& = -\frac{3}{8\pi} \sqrt{(2\ell-1)(2\ell+3)} \ell (\ell+1) m
\cos\theta\, \sin\theta\, e^{im\varphi},
\end{aligned}\\
    \label{eq:rmatac}
& \begin{aligned} 
\lefteqn{
\sum_{\ell_z=-\ell+1}^{\ell-1}
Y_{(\ell+1)(\ell_z+2m)}(\verp{}) Y_{(\ell-1)\ell_z}(\verp{})^*
\sqrt{\frac{(\ell+m\ell_z+3)!}{(\ell+m\ell_z-1)!}} = }
\hspace{45ex}\\
& = \frac{3}{8\pi} \sqrt{(2\ell-1)(2\ell+3)} \ell (\ell+1)
(\sin\theta)^2 e^{2im\varphi}~,
    \end{aligned}
  \end{align}
\end{subequations}
with $m=\pm1$ both in (\ref{eq:rmatab}) and (\ref{eq:rmatac}).
Equation (\ref{eq:rmataa}) is well known \cite{var88}, equations
(\ref{eq:rmatab}), (\ref{eq:rmatac}) extend it to $\ell'_z=\ell_z\pm1$
and $\ell_z\pm 2$, resp., and (\ref{eq:rmat7}) to
$\verpp{}\neq\verp{}$.

Setting $n=1=t$ in (\ref{eq:mix6}) we obtain
\begin{equation}
  \label{eq:mix7}
  \begin{split}
\sum_{\ell_z=-\ell}^\ell Y_{(\ell+1)(\ell_z+m)}(\verp{})
Y_{\ell\ell_z}(\verp{})^* \ell_z
\sqrt{(\ell+1+m\ell_z)(\ell+2+m\ell_z)} =
\hspace{35ex}\\
=-\frac{1}{8\pi} \sqrt{2\ell+1} \sqrt{2\ell+3}\, \ell (\ell+1) 
\sin\theta e^{i m \varphi}~,
  \end{split}
\end{equation}
with $m=\pm1$, whereas the case $m=0$ leads to a trivial identity.
For $n=1$, $t=2$ with $m=0$, $\pm1$, from (\ref{eq:mix6}) we have
\begin{subequations}
\label{eq:mix8}
\begin{align}
\label{eq:mix8a}    
&\begin{aligned}
\lefteqn{
\sum_{\ell_z=-\ell}^\ell Y_{(\ell+1)(\ell_z+m)}(\verp{})
Y_{\ell\ell_z}(\verp{})^* {\ell_z}^2
\sqrt{(\ell+1+m\ell_z)(\ell+2+m\ell_z)} =}
\hspace{25ex}\\
&=\frac{m}{32\pi} \sqrt{(2\ell+1)(2\ell+3)}\, \ell (\ell+1) \left(
(\ell+2) \sin(3\theta) - (3\ell+2) \sin\theta
\right) e^{im\varphi},
\end{aligned}\\
\label{eq:mix8b}    
&\sum_{\ell_z=-\ell}^\ell Y_{(\ell+1)\ell_z}(\verp{})
Y_{\ell\ell_z}(\verp{})^* {\ell_z}^2
\sqrt{(\ell+1)^2-\ell_z^2} =
\frac{1}{8\pi} \sqrt{(2\ell+1)(2\ell+3)}\, 
\frac{(\ell+2)!}{(\ell-1)!} \cos\theta (\sin\theta)^2~.
\end{align}
\end{subequations}
In (\ref{eq:mix7}), (\ref{eq:mix8}) we specialized the general form
(\ref{eq:mix6}) to $\ell_1=\ell+1$, the case $\ell_1=\ell-1$ following
from them by redefinition of parameters.  Those results are to be
compared with (\ref{eq:rmat9}).  Similarly, for $n=2$, $t=1$ from
(\ref{eq:mix6}) we get 
\begin{subequations}
  \label{eq:mix9}
\begin{align}
  \label{eq:mix9a}
&\begin{aligned}
\lefteqn{
\sum_{\ell_z=-\ell+1}^{\ell-1} Y_{(\ell+1)(\ell_z+m)}(\verp{})
Y_{(\ell-1)\ell_z}(\verp{})^* \ell_z
\sqrt{\ell^2-\ell_z^2} \sqrt{(\ell+1+m\ell_z)(\ell+2+m\ell_z)} 
= } \hspace{45ex}\\  
& =-\frac{1}{8\pi} \sqrt{(2\ell-1)(2\ell+3)} \frac{(\ell+1)!}{(\ell-1)!}
\cos\theta \sin\theta e^{i m \varphi},  
\end{aligned}\\
  \label{eq:mix9b}
&\begin{aligned}
\lefteqn{
\sum_{\ell_z=-\ell+1}^{\ell-1} Y_{(\ell+1)(\ell_z+2m)}(\verp{})
Y_{(\ell-1)\ell_z}(\verp{})^* \ell_z 
\sqrt{\frac{(\ell+m\ell_z+3)!}{(\ell+m\ell_z-1)!}}
= } \hspace{45ex}\\  
&=\frac{m}{4\pi} \sqrt{(2\ell-1)(2\ell+3)} \frac{(\ell+1)!}{(\ell-1)!}
(\sin\theta)^2 e^{2i m \varphi},  
\end{aligned}
\end{align}
\end{subequations}
with $m=\pm1$.  Finally, for $n=2=t$ (\ref{eq:mix6}) yields
\begin{subequations}
  \label{eq:mixa}
\begin{align}
  \label{eq:mixaa}
&\begin{aligned}
\lefteqn{
\sum_{\ell_z=-\ell+1}^{\ell-1} Y_{(\ell+1)\ell_z}(\verp{})
Y_{(\ell-1)\ell_z}(\verp{})^* {\ell_z}^2
\sqrt{\ell^2-\ell_z^2} \sqrt{((\ell+1)^2 - \ell_z^2)} 
= } \hspace{23ex}\\  
& =\frac{1}{64\pi} \sqrt{(2\ell-1)(2\ell+3)} \frac{(\ell+2)!}{(\ell-2)!}
(\sin\theta)^2 (5\cos(2\theta)+3),
\end{aligned}\\  
&\begin{aligned}
\lefteqn{
\sum_{\ell_z=-\ell+1}^{\ell-1} Y_{(\ell+1)(\ell_z+m)}(\verp{})
Y_{(\ell-1)\ell_z}(\verp{})^* {\ell_z}^2
\sqrt{\ell^2-\ell_z^2} \sqrt{(\ell+1+m\ell_z)(\ell+2+m\ell_z)} 
= } \hspace{23ex}\\  
& =\frac{m}{32\pi} \sqrt{(2\ell-1)(2\ell+3)} \frac{(\ell+1)!}{(\ell-1)!}
\cos\theta \sin\theta (5(\ell+2)(\cos\theta)^2 - 5\ell -6) e^{i m \varphi},  
\end{aligned}\\
&\begin{aligned}
\lefteqn{
\sum_{\ell_z=-\ell+1}^{\ell-1} Y_{(\ell+1)(\ell_z+2m)}(\verp{})
Y_{(\ell-1)\ell_z}(\verp{})^* {\ell_z}^2
\sqrt{\frac{(\ell+3+m\ell_z)!}{(\ell-1+m\ell_z)!}}
= } \hspace{23ex}\\  
& =-\frac{1}{32\pi} \sqrt{(2\ell-1)(2\ell+3)} \frac{(\ell+1)!}{(\ell-1)!}
(\sin\theta)^2 (5(\ell+2)(\cos\theta)^2 - 7\ell+2) e^{2 i m \varphi},  
\end{aligned}
\end{align}
\end{subequations}
where in the last two equalitites $m=\pm1$.  Many other particular
addition theorems can be derived from the general form
(\ref{eq:mix6}).  We will refrain from further discussing them here.
\end{document}